\documentclass[useAMS,usenatbib, a4, proof]{mn2e}
\usepackage{graphicx}
\usepackage{times}
\usepackage{amssymb}
\usepackage[fleqn]{amsmath}
\usepackage[draft]{hyperref}
\usepackage{wasysym}
\usepackage{textcomp}

\newcommand{\beq}{\begin{equation}}
\newcommand{\eeq}{\end{equation}}

\newcommand{\lsim}{\ \raise
-2.truept\hbox{\rlap{\hbox{$\sim$}}\raise5.truept\hbox{$<$}\ }}
\newcommand{\gsim}{\ \raise
-2.truept\hbox{\rlap{\hbox{$\sim$}}\raise5.truept\hbox{$>$}\ }}
\newcommand{\simsim}{\ \raise
-2.truept\hbox{\rlap{\hbox{$\sim$}}\raise5.truept\hbox{$\sim$}\ }}

\def\gtorder{\mathrel{\raise.3ex\hbox{$>$}\mkern-14mu
                \lower0.6ex\hbox{$\sim$}}}
\def\ltorder{\mathrel{\raise.3ex\hbox{$<$}\mkern-14mu
                \lower0.6ex\hbox{$\sim$}}}

\def\arcsec{\hbox{$^{\prime\prime}$}}
\def\solar{\mbox{$_{\normalsize\odot}$}}

\def\deg{\hbox{$^\circ$}}

\def\aj{AJ}                   
\def\araa{ARA\&A}             
\def\apj{ApJ}                 
\def\apjl{ApJL}                
\def\apjs{ApJS}               
           
\def\apss{Ap\&SS}             
\def\aap{A\&A}

\def\fcp{Fundam. Cosm. Phys.}    
             
\def\memras{MmRAS}            
\def\mnras{MNRAS}

\def\pasp{PASP}

\def\fcp{Fund.~Cosmic~Phys.}

\def\jgr{J.~Geophys.~Res.}

\def\physscr{Phys.~Scr.}

\title[Hierarchical star formation across the spiral galaxy NGC\,1566]{Hierarchical star formation across the grand design spiral NGC1566}

\author[D. A. Gouliermis et al.]
  {Dimitrios A. Gouliermis,$^{1,2,}$\thanks{E-mail: dgoulier@mpia.de, gouliermis@uni-heidelberg.de}
    Bruce G. Elmegreen,$^{3}$
   Debra M. Elmegreen,$^{4}$
 \newauthor
   Daniela Calzetti,$^{5}$
   Michele Cignoni,$^{6}$
  John S. Gallagher III,$^{7}$
Robert C. Kennicutt,$^{8}$
 \newauthor
 Ralf S. Klessen,$^{1}$   
  Elena Sabbi,$^{9}$ 
  David Thilker,$^{10}$ 
  Leonardo Ubeda,$^{9}$  
  Alessandra Aloisi,$^{9}$ 
 \newauthor
   Angela Adamo,$^{11}$
   David O. Cook,$^{12,13}$   
   Daniel Dale,$^{13}$   
      Kathryn Grasha,$^{5}$   
Eva K. Grebel,$^{14}$
 \newauthor   
   Kelsey E. Johnson,$^{15}$
   Elena Sacchi,$^{16,17}$
   Fayezeh Shabani,$^{14}$
   Linda\,J. Smith,$^{18}$
 \newauthor   
    Aida Wofford$^{19,20}$
 \\
  \\
      $~~^1$Zentrum f\"ur Astronomie der Universit\"at Heidelberg, Institut f\"ur Theoretische Astrophysik, Albert-Ueberle-Str.\,2, 69120 Heidelberg, Germany\\
      $~~^2$Max Planck Institute for Astronomy,  K\"{o}nigstuhl\,17, 69117 Heidelberg, Germany \\
      $~~^3$IBM Research Division, T.J. Watson Research Center, Yorktown Hts., NY 10598, USA \\
      $~~^4$Vassar College, Dept. of Physics and Astronomy, Poughkeepsie, NY 12604, USA \\
      $~~^5$Department of Astronomy, University of Massachusetts -- Amherst, Amherst, MA 01003, USA \\
      $~~^6$Department of Physics, University of Pisa, Largo Pontecorvo 3, 56127 Pisa, Italy \\
      $~~^7$Department of Astronomy, University of Wisconsin-Madison, WI 53706, USA \\
      $~~^8$Institute of Astronomy, University of Cambridge, Madingley Road, Cambridge CB3 0HA, United Kingdom \\
      $~~^9$Space Telescope Science Institute, 3700 San Martin Drive, Baltimore, MD 21218, USA \\
      $^{10}$Department of Physics and Astronomy, Johns Hopkins University, 3701 San Martin Drive, Baltimore, MD 21218, USA\\
      $^{11}$Department of Astronomy, Oskar Klein Centre, Stockholm University, AlbaNova University Centre, SE-106 91 Stockholm, Sweden \\
      $^{12}$California Institute of Technology, Pasadena, CA 91125, USA\\
      $^{13}$Department of Physics and Astronomy, University of Wyoming, Laramie, WY 82071, USA \\
      $^{14}$Astronomisches Rechen-Institut, Zentrum f\"ur Astronomie der Universit\"at Heidelberg, M\"onchhofstr. 12-14, 69120 Heidelberg, Germany \\
      $^{15}$Department of Astronomy, University of Virginia, P.O. Box 400325, Charlottesville, VA 22904-4325, USA \\  
      $^{16}$Department of Physics and Astronomy, Bologna University, Viale Berti Pichat 6/2, 40127 Bologna, Italy \\
      $^{17}$INAF-Osservatorio Astronomico di Bologna, Via Ranzani 1, 40127 Bologna, Italy \\
      $^{18}$European Space Agency and Space Telescope Science Institute, 3700 San Martin Drive, Baltimore, MD 21218, USA  \\
      $^{19}$Sorbonne Universit\'{e}s, UPMC-CNRS, UMR7095, Institut dÕAstrophysique de Paris, F-75014 Paris, France \\
      $^{20}$Instituto de Astronom\'{i}a, Universidad Nacional Aut\'{o}noma de M\'{e}xico, Unidad Acad\'{e}mica en Ensenada, Ensenada 22860, Mexico
  }

\begin{document}

\date{Accepted 2017 February 17. Received 2017 February 09 ; in original form 2017 January 11}

\pagerange{\pageref{firstpage}--\pageref{lastpage}} \pubyear{2015}

\maketitle

\label{firstpage}


\begin{abstract}
We investigate how star formation is spatially organized in the grand-design spiral NGC\,1566 from deep HST photometry with the {\sl Legacy ExtraGalactic UV Survey} (LEGUS). Our contour-based clustering analysis reveals 890 distinct stellar conglomerations at various levels of significance. These star-forming complexes are organized in a hierarchical fashion with the larger congregations consisting of smaller structures, which themselves fragment into even smaller and more compact stellar groupings. Their size distribution, covering a wide range in length-scales, shows a power-law as expected from scale-free processes. We explain this shape with a simple ``fragmentation and enrichment'' model. The hierarchical morphology of the complexes is confirmed by their mass--size relation which can be represented by a power-law with a fractional exponent, analogous to that determined for fractal molecular clouds. The surface stellar density distribution of the complexes shows a log-normal shape similar to that for supersonic non-gravitating turbulent gas. Between 50 and 65 per cent of the recently-formed stars, as well as about 90 per cent of the young star clusters, are found inside the stellar complexes, located along the spiral arms. We find an age-difference between young stars inside the complexes and those in their direct vicinity in the arms of at least 10 Myr. This timescale may relate to the minimum time for stellar evaporation, although we cannot exclude the in situ formation of stars. As expected, star formation preferentially occurs in spiral arms. Our findings reveal turbulent-driven hierarchical star formation along the arms of a grand-design galaxy.
\end{abstract}




\begin{figure*}
\centering
\includegraphics[width=1.\textwidth]{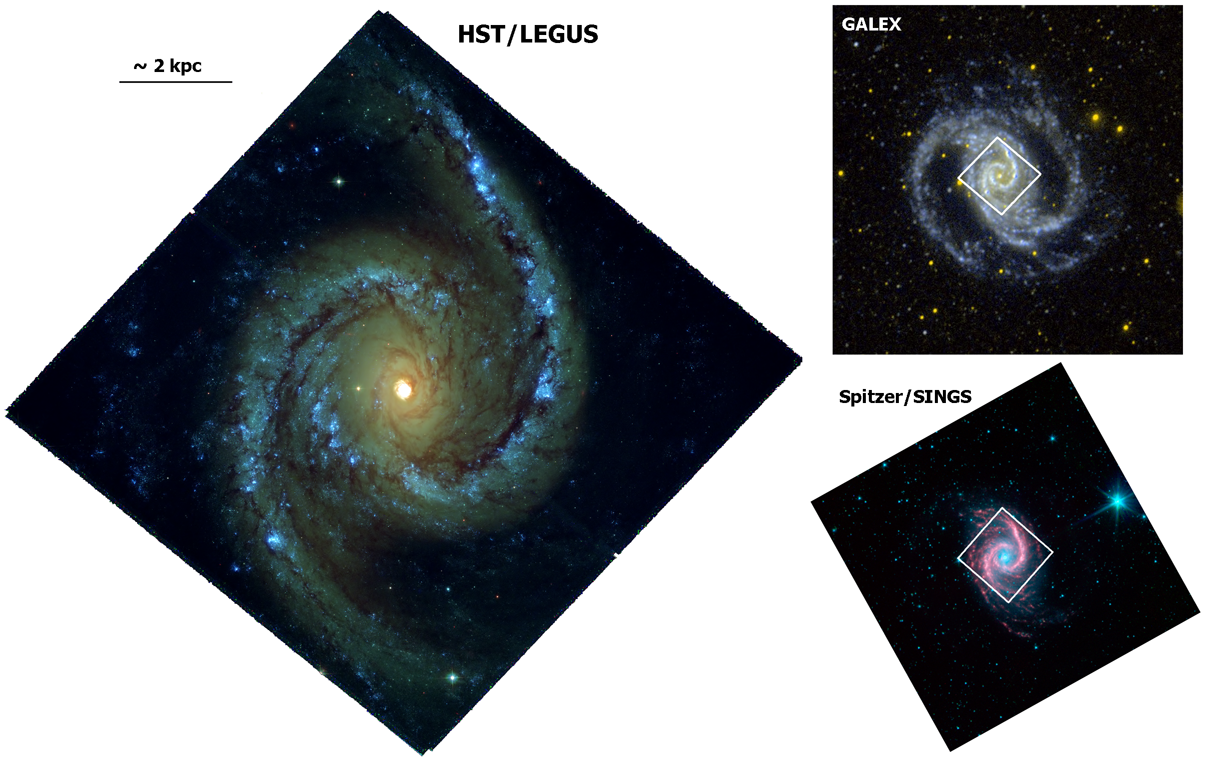} 
\caption{{\em Left:} Colour composite image of the observed WFC3 field-of-view of NGC\,1566, constructed from LEGUS imaging in the filters F336W (blue), F438W (green) and F814W (red). 
{\em Right:} GALEX two-colour (far-UV and near-UV) and Spitzer/SINGS \citep[][blue: 3.6\,$\mu$m, red: 8\,$\mu$m]{Kennicutt2003} images of the whole extent of the galaxy with the LEGUS WFC3 footprint overdrawn.  In all images North is up and East is left.
 \label{fig:ngc1566ima}}
\end{figure*}


\section{Introduction}\label{s:intro}


Star formation, the process that converts gas into stars, is a key mechanism for the formation and evolution of galaxies.
Star Formation across the disk of a spiral galaxy is in principle governed by three factors: (1) The molecular gas reservoir of the galaxy and its molecular clouds \citep{2004ApJ...606..271G}, 
(2) the dynamics and kinematics of the galactic disk \citep{2011EAS....51...19E}, and (3) the star formation efficiency and rate at various scales \citep{2012ARA&A..50..531K}. These properties, tightly dependent upon each other, determine the gravitational self-binding and stellar and gas content of newly-born star clusters and stellar associations, as well as their conspicuous structures. It has long been known that large stellar structures, named {\em stellar complexes}, are the prominent signposts of star formation in galactic disks \citep[e.g.,][]{1964ApJS....9...65V, 1989ASPRv...7..107E,2014ApJ...780...32E}. 
These stellar structures trace star formation over several orders of magnitude in length-scales, and their characteristics relate to both the global galactic properties (dynamics, gas reservoir) and local environmental conditions (turbulent cascade, feedback) that regulate star formation \citep[see, e.g.,][]{MacLowKlessen2004}. 

Star-forming complexes -- i.e., the stellar nurseries at scales equivalent to giant molecular clouds --  are usually structured 
in a hierarchical fashion, by hosting smaller and denser stellar associations and aggregates, which themselves are 
 sub-structured into more compact clusters \citep[e.g.,][]{efremov-elmegreen-98, elmegreen-ppiv}. Therefore, analysing the demographics of the stellar complexes of galaxies provides a new way to  understand how star formation is organized across galactic disks, and how the impressive spiral star-forming pattern, seen in these galaxies in UV light, is built up.  High resolution, sensitivity and wide-area coverage are critical for the identification of stars and stellar systems at various length-scales in galaxies 
in the extended Milky Way neighbourhood, and the {\sl Hubble Space Telescope} (HST) with its unique UV-sensitivity, is the only telescope that meets all three requirements.  In light of these requirements, in the HST {\sl Legacy ExtraGalactic UV Survey}\footnote{\href{https://legus.stsci.edu/}{https://legus.stsci.edu/}} \citep[LEGUS;][]{calzetti15} we performed panchromatic imaging of 50 star-forming Local Volume galaxies. 
The program focuses on the investigation of star formation and its relation with galactic environment. 

In this study we present our second detailed statistical analysis of galactic-scale star formation in disk galaxies from LEGUS-resolved young stellar populations. In our first proof-of-concept study we demonstrated that star formation follows a hierarchical morphology in the ring galaxy NGC\,6503 \citep[][]{Gouliermis2015}. We evaluated the scale-free star formation pattern through the population demographics of the stellar complexes and the distribution of massive blue stars along the star-forming ring of the galaxy. In the present investigation we focus on the star-forming complexes population and their structural and physical parameters in the spiral NGC\,1566 (Fig.\,\ref{fig:ngc1566ima}). This galaxy is referred in the literature as a {\em grand-design} spiral that demonstrates an elaborate structure with its two sets of bi-symmetric spirals (GALEX image in Fig.\,\ref{fig:ngc1566ima}). The bar and the spiral arms of NGC\,1566 inside corotation (covered by the Hubble image of Fig.\,\ref{fig:ngc1566ima}) are structured mainly by regular orbits, with chaotic orbits playing also a role in building weak extensions of the inner spirals and in the central part of the bar \citep[][]{2013MNRAS.434.2922T}. These characteristics, as well as the low inclination and the relatively isolated environment of the galaxy, make NGC\,1566 an exceptionally interesting case.

NGC\,1566, the brightest member of the Dorado Group, is an SAB(rs)b galaxy, i.e., an intermediate-type barred spiral galaxy of 
intermediate apparent bar strength, having open, knotty arms, a small bulge, and an outer pseudo-ring made from arms that wind 
about 180\deg\ with respect to the bar ends \citep{buta15}\footnote{This classification is made from the {\em Spitzer Survey of Stellar 
Structure in Galaxies} (S$^{4}$G). According to the {\em Third Reference Catalogue of Bright Galaxies} \citep{devaucouleurs91}, the 
galaxy was previously classified as SAB(s)bc, i.e., an intermediate-type barred spiral with open, knotty spiral arms, an inner ring, and a significant bulge.}. 
The galaxy hosts a low-luminosity AGN, classified as Seyfert \citep{1961MmRAS..68...69D}, although its precise type between Seyfert 1 and 2 varies in the literature \citep[e.g.,][]{Aguero2004, 2014A&A...565A..97C}. NGC\,1566 is considered a typical example of galaxy with bar-driven spiral density 
waves \citep{2010ApJ...715L..56S}. Its strong spiral arms are found to fall in the region where bar-driving is expected (covered by our LEGUS Field-of-View; 
see Fig.\,\ref{fig:ngc1566ima}), while the additional spiral beyond $\sim$\,100\arcsec\ (see, e.g., GALEX image in Fig.\,\ref{fig:ngc1566ima}) is an 
independent pattern, as suggested by various investigators \citep[e.g.,][]{1992ASSL..178..207B, Aguero2004} .

There is no consensus in the literature about the distance of NGC\,1566, which is found to vary between 5.5 and 21.3\,Mpc. 
Distances for the galaxy are reported by \cite{Tully1988}, \cite{Mathewson1992}, \cite{Willick1997}, \cite{Theureau2007}, 
\cite{Sorce2014}, and \cite{Tully2013}. All but one measurements are based on the Tully-Fisher method, and almost all 
of them are comparable (their third quartile is $\sim$\,10.8\,Mpc). Throughout this study we adopt the NED mean distance 
of $\sim$\,10\,$\pm$\,5\,Mpc, corresponding to a distance modulus of $\sim$\,30\,$\pm$\,1\,mag. This distance is confirmed 
by our optical colour-magnitude diagram, where evolutionary models (corrected for this distance and for solar metallicity)   
reproduce well the colours of the RGB tip and several evolutionary sequences. In any case, considering the literary discrepancies 
with a factor-of-two spread in the published distance estimates, it is worth noting that the main 
results in this paper do not depend sensitively on the distance of NGC\,1566 (see discussion in Appendix\, \ref{s:dist_effect}). 

The subject we wish to address with our study of NGC\,1566 is {\em to understand galactic-scale star formation from the young stellar populations across the disk of a spiral galaxy.} In particular, in grand-design galaxies star formation takes place almost entirely along their spiral arms. Both the dynamically-driven turbulence of the disk's gaseous matter at large scales and the local conditions that favour gravitational collapse at small scales effectively shape the star formation process in the arms. While global spiral wavemodes that produce grand-design patterns have little influence on large-scale star formation 
rates, they do regulate star formation by forcing the gas into dense molecular phase in the shock fronts, and 
organizing it to follow the underlying stellar spiral \citep{2011EAS....51...19E}. 

Here, we use the most accurate stellar photometry to date 
on NGC\,1566 to understand how star formation is organized in the 
arms of grand-design galaxies. The present study addresses three specific issues: (i) The statistics and correlations of structural parameters, such as size and mass, of star-forming 
complexes in typical disk galaxies. (ii) The hierarchical nature of star formation along spiral arms, and (iii) How the clustering pattern of young stars in these
galaxies is quantified at various scales. We address these issues for NGC\,1566 with the use of the rich census of young blue stars resolved with LEGUS, through the investigation of their recent star formation as imprinted in their clustering morphology in this galaxy.


This paper is organized as follows. In Sect.\,\ref{s:datamethod} we describe the LEGUS dataset of NGC\,1566 
and its photometry. We also select the stellar samples corresponding to the young blue populations of the galaxy,
and address their spatial distribution. In the same section we perform the identification of the stellar complexes 
across the galaxy. In Sect.\,\ref{s:results} we determine the structural parameters of the identified young stellar structures, 
and we present the demographics of these parameters. We also discuss the distributions of 
the sizes and densities of the complexes, as well as the correlation of their basic parameters, namely, mass, size, density 
and crossing time. We discuss our results in terms of how star formation is organized across the spiral arms of NGC\,1566. 
The fraction of young stars in the arms in comparison of the interarm regions, and the implications of their difference on 
how star formation proceeds in the galaxy are discussed in Sect.\,\ref{s:discussion}. We summarize our findings in 
Sect.\,\ref{s:summa}.



\section{Data and Identification Method}\label{s:datamethod}

\subsection{Observations and Photometry}\label{s:obs}


The HST extragalactic panchromatic stellar survey LEGUS has mapped 50 star-forming galaxies 
in the Local Volume with emphasis on UV-related astronomical research. Images 
of the galaxies have being collected with WFC3 and ACS in parallel in the coverage from 
the near-UV to the I-band. Descriptions of the survey, its scientific objectives and the data 
reduction process are given by \cite{calzetti15}. The images of NGC\,1566 presented in this 
analysis were obtained with WFC3 in the filters F275W, F336W, F438W, F555W and F814W 
(equivalent to NUV, U, B, V, and I respectively). 

A pixel-based correction for charge-transfer efficiency 
(CTE) degradation using STScI tools was performed on the images before their processing 
with {\sc astrodrizzle} and prior to their photometry. Stellar photometry was performed with the 
point-spread function (PSF) fitting package {\sc dolphot} \citep[e.g.,][]{dolphin00}. The images 
were first prepared for masking defects and splitting  the multi-image STScI FITS into 
a single FITS file per chip with {\sc dolphot} packages {\sc acsmask} and {\sc splitgroups}.  The instrumental magnitudes were calibrated 
to the VEGAMAG scale based on the zeropoints provided on the WFC3 web-page\footnote{\url{http://www.stsci.edu/hst/wfc3}}. 
The detailed stellar photometric process applied for LEGUS will be described in a dedicated paper by Sabbi et al. (in preparation).


\begin{figure}
\centering
\includegraphics[width=\columnwidth]{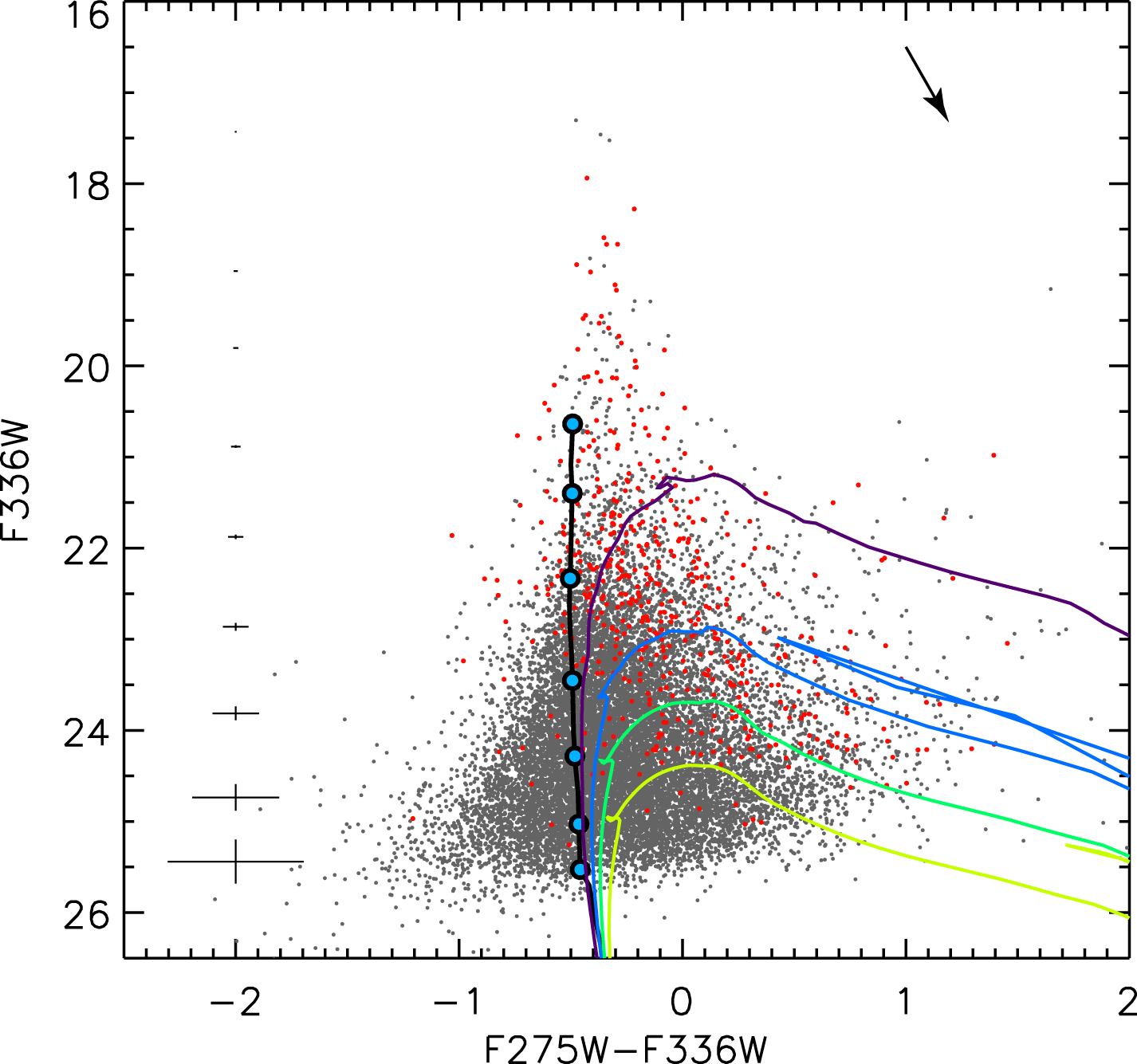} 
\caption{Colour-magnitude diagram of stars identified in NGC\,1566 with the best 
photometric quality in the LEGUS filter-pair F275W, F336W. 
This sample represents the recently-formed massive stellar population of NGC\,1566. 
The zero-age main-sequence from the Padova grid of evolutionary models for 
solar metallicity, corrected for an extinction of $A_V =$\,0.55\,mag (indicated by the 
black arrow) is overlaid with indicative positions for stars with masses 15, 20, 30, 50, 100, 
150 and 300 M$_{\odot}$. Isochrones of ages 5, 10, 15 and 20\,Myr from the same family 
of models are also plotted with various colours.  Typical photometric uncertainties in both magnitudes and colours are 
shown on the left of the CMD. Red symbols correspond to star clusters identified also by 
our stellar photometry (see Sect.\,\ref{s:obs}).
\label{fig:cmds}}
\end{figure}

\begin{figure}
\centering
\includegraphics[width=1.\columnwidth]{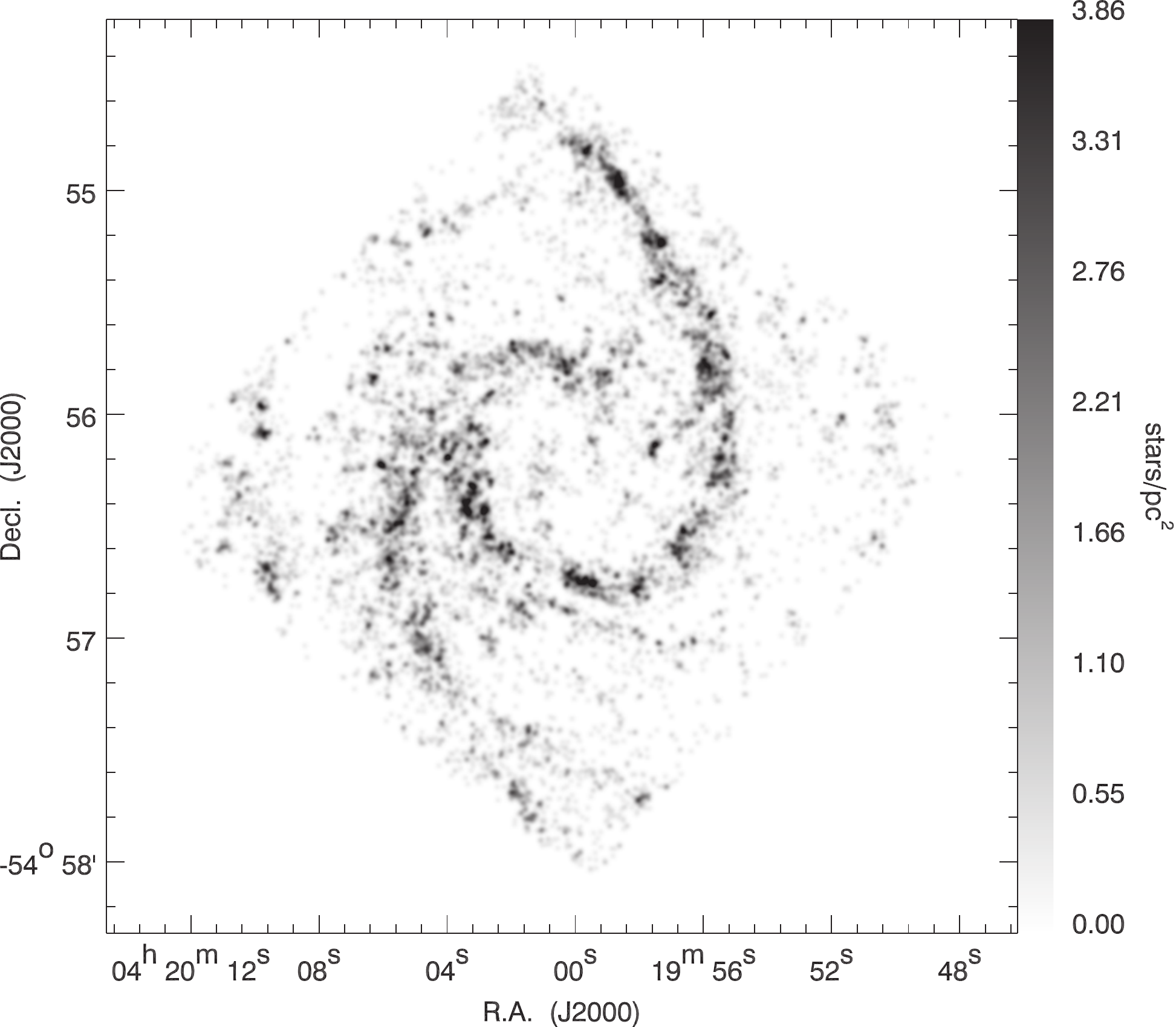} 
\caption{Surface stellar density maps constructed with the {\em Kernel Density Estimation} method with a kernel of FWHM\,$\sim$\,67\,pc, 
for stars identified with the best photometry in the filter pairs F275W, F336W (bright blue stellar sample). This surface map tracks extremely 
well the two symmetric  spiral arms of NGC\,1566. Our clustering analysis reveals individual star-forming structures across the arms, which we investigate 
in order to access the large-scale progression of star formation across the galaxy. The map is shown in linear scale. The grey-scale bar at the right 
corresponds to the stellar surface density in stars/pc$^2$.  This map demonstrates that star formation occurs mainly along the bi-symmetric spiral structures, 
corroborating the density wave-driven spiral pattern of the galaxy, recently confirmed by Shabani et al. (in preparation) from the star cluster age sequence across the arms of NGC\,1566.\label{fig:kde_maps}}
\end{figure}

The photometry with {\sc dolphot} returns several fit-quality parameters for each of the detected sources. 
The most probable stars have the object type parameter with a value of unity, while  sources too faint for PSF 
determination and non-stellar objects have  {\sc type}\,$>1$.  The photometry file also includes the crowding parameter,
which is a measure of how much brighter the star would have been measured had nearby stars not been fit simultaneously. 
For an isolated star this parameter has a value of zero. A perfectly-fit star has a sharpness parameter equal to zero; this 
parameter will be positive for a star that is too sharp, and negative for a star that is too broad. More details about the 
quality parameters are given in {\sc dolphot} documentation\footnote{Available at \href{http://americano.dolphinsim.com/dolphot/}{http://americano.dolphinsim.com/dolphot/}}.
We determine the {\em best photometrically defined} stellar samples in terms of these quality parameters
by applying a set of selection criteria for the identified stellar sources: 
\begin{itemize}
\item[] {\sc dolphot} type of the source, {\sc type}\,$=1$
\item[] Crowding of the source in each of the filters,  {\sc crowd}\,$< 2$ 
\item[] Sharpness of the source squared in each filter, {\sc sharp}$^2< 0.3$ 
\item[] Signal-to-noise ratio in each filter, {\sc snr}\,$> 5$ 
\end{itemize}

We separate from the stellar sources with the most reliable photometry the sample that includes stars found in the 
filter pair (F275W, F336W), which corresponds to the younger bright blue population 
of the galaxy. The star formation analysis we present in the following sections is based on this 
stellar sample, which includes 14,928 sources (we refer to it as the `blue stellar sample').
A second stellar sample, covering stars with photometric measurements in the 
filter pair (F555W, F814W), but not in the (F275W, F336W) filter pair, comprising 18,050 stars,
was also selected. This sample, corresponding to the evolved young stellar populations with ages up to $\sim$\,80\,Myr,
will be discussed in another paper dedicated to the time-evolution of spiral structure.

The colour-magnitude diagram (CMD), of the blue stellar sample is shown in Fig.\,\ref{fig:cmds}.  
Stellar evolutionary isochrones from the Padova grid of models (\citealt{chen15}; see also \citealt{marigo08, girardi10, bressan12}) are also shown. 
The models are corrected for an indicative extinction of $A_V=0.55$\,mag (determined with isochrone fitting), assuming an extinction coefficient $R_V=3.1$ and the reddening law of \cite{Fitzpatrick1999}, recalibrated for the WFC3 photometric system by \cite{Schlafly2010}.
From this CMD it is shown that the star-forming populations in NGC\,1566 correspond to ages \lsim\,20\,Myr. The youthfulness of these populations 
is also demonstrated  by the zero-age main-sequence (ZAMS), constructed from the same models family (thick black line). In the plot  
we also mark the positions of ZAMS stars for masses starting at 15\,M{\solar} and reaching the theoretical extreme 
of 300\,M{\solar}.  According to the ZAMS model our photometric detection limit in the blue CMD corresponds to stars with $\sim$\,15\,M{\solar}. 
It should be noted that the luminosity mismatch between the tip of the ZAMS and the brightest observed objects is possibly due to the fact that the F336W filter is not sampling entirely the flux of stars $\geq$\,200\,M{\solar}. Nevertheless, the existence of main-sequence sources far brighter than this mass limit indicates that these objects are possibly blended systems of multiple bright blue stars.

\begin{figure*}
\centering
\includegraphics[width=\textwidth]{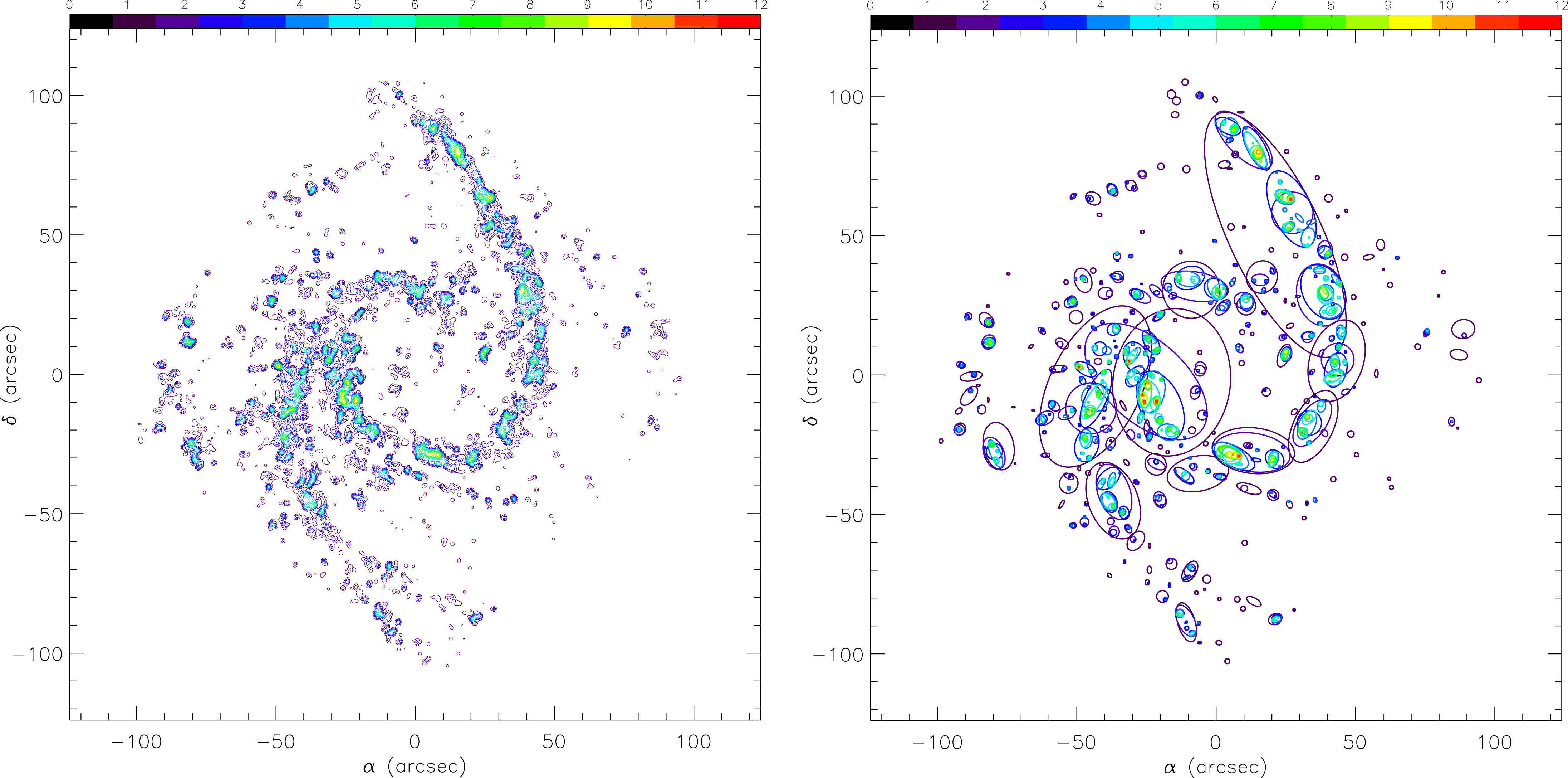} 
\caption{{\em Left panel.} Iso-density contour plot of the surface stellar density map constructed with the KDE method with a kernel of 
FWHM\,$\sim$\,67\,pc, for the blue stars identified in both filters F275W and F336W. {\em Right panel.} Chart of the identified 
stellar structures. System borders are drawn with ellipses as determined by fitting the convex-hull of the systems at the various 
significance levels, where they appear. Different colours, used for both the isopleths and the ellipses of the systems, correspond to 
different significance levels (in $\sigma$), drawn in 1$\sigma$ steps. The defined ellipses are used only for demonstrating the 
geometry of each complex and an estimate of its elongation. The structural parameters of the complexes are determined by the stellar 
sources encompassed within their boundaries, as defined by their actual isopleths (see Sect.\,\ref{s:params}).
\label{fig:contour_map}}
\end{figure*}

We performed a cross-matching (with a search radius of 0.1\arcsec) between the stellar photometric catalogue and the catalogue of the most probable star clusters identified across NGC\,1566 by the LEGUS team (cluster catalogue version PadAGB\_MWext\_04Nov15). The method used to produce LEGUS cluster catalogues is described in detail in Adamo et al. (in preparation). 
From the 677 young star clusters, with ages $\leq$\,100\,Myr,  identified  in all three considered classes\footnote{Candidate clusters were classified as {\em class 1} (compact, centrally concentrated objects, with a FWHM more extended than stellar), {\em class 2} (objects with slightly elongated density profiles and less symmetric light distribution), and {\em class 3} (less compact objects showing asymmetric profiles and multiple peaks). Details on the classification scheme are
given in Adamo et al. (in preparation).}, 518 objects (77\,per cent of the sample) were identified also by our blue stellar photometry. These 
systems are shown with red symbols in the CMD of Fig.\,\ref{fig:cmds}. The integrated magnitudes of the clusters derived with aperture photometry within
the cluster selection process are in very good agreement but systematically brighter than the corresponding PSF magnitudes derived from the stellar photometry. These sources, corresponding to the small fraction of 3\,per\,cent of the total blue population, do not influence at all our statistical analysis. Nevertheless, considering that stellar complexes comprise {\em by definition} multiple systems, associations and 
clusters, apart from individual stars, we include these sources in our treatment.

\subsection{Stellar Surface Density Maps}\label{s:method}

The spatial distribution of the bright blue stars in NGC\,1566 is shown in the stellar surface density map of 
Fig.\,\ref{fig:kde_maps}. 
This map is constructed  with the application of the {\em Kernel Density Estimation} (KDE), i.e., by 
convolving the map of detected blue sources with a Gaussian kernel. The FWHM of this kernel depends on the purpose 
of the KDE map. In the case of the map of Fig.\,\ref{fig:kde_maps}, which will be used for the 
identification of the complexes population of OB-stars in the galaxy the kernel specifies the ``resolution'' 
at which stellar structures will be revealed. For this identification the stellar density map should not be 
smooth enough to erase any fine-structure and it should not be detailed enough to introduce any significant 
noise. 

In general the `optimal' kernel size depends on the data completeness and the distance of the galaxy, and therefore 
it is best decided upon experimentation. Testing various kernel sizes for the blue stellar sample showed that a FWHM 
of $\sim$\,1.4\arcsec, corresponding to a physical scale of $\sim$\,67\,pc, is the minimum possible for the detection of 
the star-forming complexes  of the galaxy. This scale compares well to the typical size of  
OB associations in the Local Group \citep[][Table\,1]{gouliermis11} and of molecular clouds \citep[see, e.g.,][]{bolatto08} in various galaxies.  

From the KDE map of Fig.\,\ref{fig:kde_maps} it is seen that the recently-formed stellar population tracks extremely well the spiral features 
of the galaxy. We compared the blue stellar distribution against the light distribution from Spitzer/SINGS images in  8\,$\mu$m and 24\,$\mu$m \citep{Kennicutt2003}, indicators of the loci of young stars based on dust emission, in order to check whether the observed stellar distribution is affected by dust attenuation. This comparison confirmed that the blue stars are more clustered along the spiral arms of the galaxy. While MIPS 
resolution shows only the general coincidence in the distributions of stars and 24\,$\mu$m emission, the IRAC 8\,$\mu$m image traces well individual large young stellar structures seen in our density map. Most of these structures in the KDE map consist of smaller more compact structures, which themselves ``break'' into even smaller and denser ones. We discuss this hierarchical clustering behaviour in the following section. 

\begin{table*}
\centering
\caption{Demographics of the young stellar complexes, identified in NGC\,1566 at various significance levels.
In Cols. 1 and 2 the detection levels (in $\sigma$) and the number of detected structures are given.  The parameters shown include the
  minimum, average and maximum size (Cols.\,3, 4 and 5) of the complexes (see Sect.\,\ref{s:basic_params}), 
  and the average stellar surface density  (Col.\,6) of the structures
in each density level. The corresponding total stellar masses ($M_{\star}$,
Col.\,7) and UV magnitudes ($m_{\rm 275}$, Col.\,9) per density level are also given, along with the corresponding fractions of these
parameters over the total stellar mass, $M_{\star, {\rm tot}}$, and total stellar UV flux, $m_{\rm tot}$ (Cols.\,8 and 10, respectively).
Average crossing times and velocity dispersions of the systems per detection level are given in Cols. 11 and 12. The parameters for each stellar 
complex are determined by extrapolation of the observed mass function of the total stellar sample (Sect.\,\ref{s:advanced_params}).}
\label{tab:structdemo}
\begin{tabular}{rrrrrrrcccrc}
\hline
\multicolumn{1}{c}{Level} &
\multicolumn{1}{c}{$N_{\rm sys}$} &
\multicolumn{3}{c}{Size $S$ (pc)} &
\multicolumn{1}{c}{$\langle \Sigma_{\star}\rangle$} &
\multicolumn{1}{c}{$M$} &
\multicolumn{1}{c}{$f_{\star}$}&
\multicolumn{1}{c}{$m_{\rm 275}$}&
\multicolumn{1}{c}{$f_{\rm UV}$}&
\multicolumn{1}{c}{$\langle t_{\rm cr}\rangle$}&
\multicolumn{1}{c}{$\langle \sigma_{\upsilon} \rangle$} \\
\multicolumn{1}{c}{($\sigma$)} &
\multicolumn{1}{c}{} &
\multicolumn{1}{c}{$S_{\rm min}$} &
\multicolumn{1}{c}{$\langle S \rangle$} &
\multicolumn{1}{c}{$S_{\rm max}$} &
\multicolumn{1}{c}{(M{\solar}\,pc$^{-2}$)}&
\multicolumn{1}{c}{($10^{4}$\,M{\solar})}&
\multicolumn{1}{c}{$M_{\star}/M_{\star,{\rm tot}}$}&
\multicolumn{1}{c}{(mag)}&
\multicolumn{1}{c}{$s_{\rm 275}/s_{\rm tot}$}&
\multicolumn{1}{c}{(Myr)}&
\multicolumn{1}{c}{(km\,s$^{-1}$)}\\
\multicolumn{1}{c}{(1)} &
\multicolumn{1}{c}{(2)} &
\multicolumn{1}{c}{(3)} &
\multicolumn{1}{c}{(4)}&
\multicolumn{1}{c}{(5)}&
\multicolumn{1}{c}{(6)}&
\multicolumn{1}{c}{(7)}&
\multicolumn{1}{c}{(8)}&
\multicolumn{1}{c}{(9)}&
\multicolumn{1}{c}{(10)}&
\multicolumn{1}{c}{(11)}&
\multicolumn{1}{c}{(12)}\\
\hline 
   1 &  172 &     108.7 &     241.0 &    1855.8 &     0.19 $\pm$    0.02 &     354.9 &    0.539 &   12.60 &    0.836 &   405.58 &      0.5 $\pm$     0.2 \\
   2 &  113 &     109.2 &     233.9 &    1390.4 &     0.30 $\pm$    0.03 &     268.6 &    0.408 &   12.90 &    0.632 &   323.34 &      0.6 $\pm$     0.2 \\
   3 &  161 &      60.8 &     161.9 &     774.4 &     0.44 $\pm$    0.08 &     222.6 &    0.338 &   13.11 &    0.524 &   226.34 &      0.6 $\pm$     0.2 \\
   4 &  134 &      45.1 &     150.9 &     540.0 &     0.55 $\pm$    0.16 &     165.3 &    0.251 &   13.43 &    0.389 &   201.49 &      0.7 $\pm$     0.2 \\
   5 &   87 &      69.5 &     152.8 &     457.7 &     0.58 $\pm$    0.09 &     112.8 &    0.171 &   13.85 &    0.265 &   194.30 &      0.7 $\pm$     0.2 \\
   6 &   81 &      41.9 &     121.5 &     362.9 &     0.71 $\pm$    0.17 &      78.6 &    0.119 &   14.24 &    0.185 &   158.78 &      0.7 $\pm$     0.2 \\
   7 &   55 &      38.1 &     109.8 &     306.3 &     0.82 $\pm$    0.22 &      49.0 &    0.075 &   14.75 &    0.115 &   141.66 &      0.7 $\pm$     0.2 \\
   8 &   37 &      48.6 &     102.4 &     258.2 &     0.84 $\pm$    0.16 &      29.5 &    0.045 &   15.30 &    0.070 &   133.42 &      0.7 $\pm$     0.1 \\
   9 &   22 &      27.7 &      91.9 &     192.6 &     0.98 $\pm$    0.37 &      15.2 &    0.023 &   16.02 &    0.036 &   119.90 &      0.7 $\pm$     0.1 \\
  10 &   12 &      32.3 &      86.5 &     151.1 &     1.15 $\pm$    0.39 &       8.0 &    0.012 &   16.72 &    0.019 &   108.32 &      0.8 $\pm$     0.1 \\
  11 &   10 &      51.4 &      69.3 &      86.0 &     1.19 $\pm$    0.34 &       4.4 &    0.007 &   17.37 &    0.010 &    95.09 &      0.7 $\pm$     0.1 \\
  12 &    6 &      32.5 &      48.7 &      62.0 &     1.60 $\pm$    0.69 &       1.7 &    0.003 &   18.40 &    0.004 &    70.88 &      0.7 $\pm$     0.1 \\
\hline
\end{tabular}
\end{table*}

\subsection{Detection of Stellar Complexes}\label{s:detection}

The KDE stellar density map of Fig.\,\ref{fig:kde_maps} 
is a statistical significance map, corresponding to the two dimensional probability function of the clustering 
of the stars in the blue sample. We apply a contour-based clustering analysis technique on this map to 
identify all star-forming complexes in NGC\,1566. 
Individual stellar complexes are identified as distinct stellar over-densities at various levels of significance, 
defined in $\sigma$ above the background density ($\sigma$ being the standard deviation of the map). 
The values of the background density of the map and its standard deviation are $\sim 2 \cdot 10^{-4}$\,arcsec$^{-2}$ and 
$\sim 5 \cdot 10^{-4}$\,arcsec$^{-2}$ respectively.  Each structure 
is defined by its closed iso-density contour line at the significance level of its detection. We start the identification at the 
 level of 1$\sigma$ and we repeat the detection process at higher density levels, in steps 
of 1$\sigma$. We construct thus a survey of asymmetric large young stellar concentrations that covers the complete dynamic range in 
stellar density \citep[see, e.g.,][for original implementations of the method]{gouliermis00, gouliermis10}.  Smaller, 
and more compact stellar concentrations are found systematically within the borders of larger and looser ones, providing clear 
evidence of hierarchy in the distribution of the bright blue stars in NGC\,1566. 



The surface density map of the blue stellar population is depicted in Fig.\,\ref{fig:contour_map} (left panel) 
as a contour map with {\em isopleths}\footnote{An ``isopleth'' defines a line on the map that connects points having 
equal surface stellar density; Origin from the Greek {\em ``isopl\={e}th\={e}s''}, equal in number.} drawn with different colours according to their corresponding significance levels. Each identified stellar complex is represented by an ellipse in the chart of the survey shown in the right panel of Fig.\,\ref{fig:contour_map}. In both plots the level-colour correspondence is indicated by the colour-bar at the top. These maps demonstrate more vividly the partitioning of the spiral arms of the galaxy into stellar structures, which are revealed at various density (significance) levels. 
In total 949 stellar structures are revealed with our clustering analysis, corresponding to various density levels up to the highest level of 12$\sigma$. The determination of the effective radius, $r_{\rm eff}$, of each stellar complex and its size  $S$, and the evaluation of the ellipse that best represents its morphology are discussed in Sect.\,\ref{s:basic_params}. It should be noted that the ellipses are determined only for 
the demonstration of the geometry of the structures and the estimation of their elongation. The structural parameters of the complexes are determined by the actual isopleths, which define their borders, as described in Sect.\,\ref{s:basic_params}.

An important parameter considered in compiling this survey of stellar complexes is the minimum number of stars counted within the borders of each detected structure. 
In order to eliminate the contamination of our survey by random stellar congregations, so-called asterisms, we confine our catalogue to 
structures identified with at least 5 members  \citep[e.g.,][]{bastian07}, reducing the number of identified structures to 890.  
The appearance of structures in at least two consecutive significance levels provides confidence that these are real stellar 
concentrations -- this condition was used as an identification criterion in \cite{Gouliermis2015}. In the present catalogue of complexes 
there are 59 structures identified at the 1$\sigma$ level, for which there are no counterparts at any other higher level. While this fact 
provides the ground for disregarding these objects as spurious detections, their positions coincide with prominent faint brightness 
patches in the UV and U images of NGC\,1566 and in accordance with the spiral features of the galaxy. Therefore, we consider
these faint loose structures in our further analysis as real stellar complexes. These systems are indicated by the 
small single 1$\sigma$ ellipses (black lines) in the chart of Fig.\,\ref{fig:contour_map} (right panel). 



\begin{figure}
\centerline{\includegraphics[clip=true,width=\columnwidth]{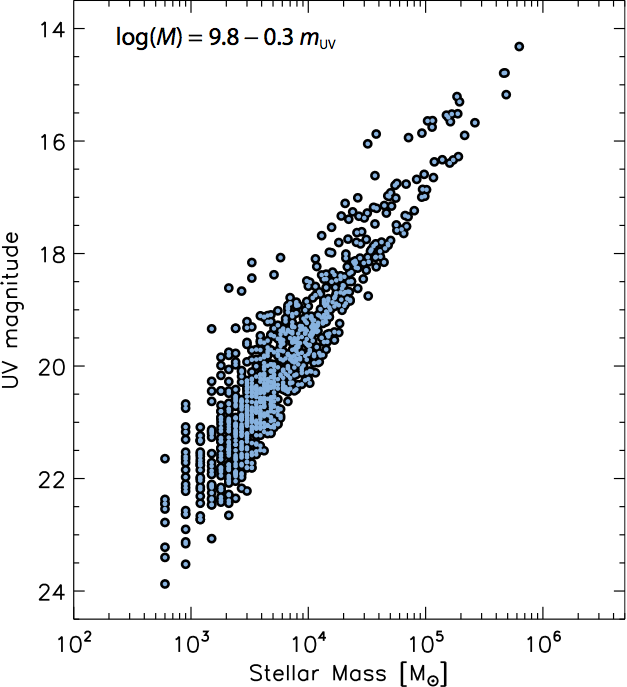}}
\caption{Relation between total stellar masses, determined from the observed stellar MF extrapolation, and observed UV brightness (not corrected for extinction) in magnitudes in the filter F275W for the stellar complexes in NGC\,1566. This relation (with a correlation coefficient $\simeq$\,0.88) 
shows that the derived stellar masses correlate quite well with the UV luminosities in the relevant structures.
\label{f:masslight}}
\end{figure}

\section{Results}\label{s:results}

\subsection{Parameter Determination}\label{s:params}

We derive structural parameters for the stellar complexes on the basis of their observed stellar masses, using the number of blue stars enclosed within the borders of each structure, and their sizes. The stellar mass estimated for each structure allows for the determination of its surface stellar mass density. We assess the dynamical status of the detected complexes with the evaluation of their crossing times and velocity dispersions, which also depend on stellar mass. The total stellar mass encompassed in each stellar complex is determined in terms of 
extrapolation of the mass function of the total stellar sample, as described in the following section. The calculated masses, though, suffer from 
observational constraints such as photometric incompleteness and evolutionary effects, which affect the derived stellar masses and ages. They cannot, thus, be taken at face value, but they have important comparative value to our analysis of the distributions of the derived parameters and their correlations, presented later in this study. Based on the observed blue CMD we assume an age for all systems of at most 20\,Myr. The young age of the detected complexes is confirmed by their spatial coincidence with bright H\,{\sc ii} regions identified on ground-based H$\alpha$  narrow-band plates \citep[][]{1982A&A...114....7C}.

\subsubsection{Basic (first generation) parameters}\label{s:basic_params}

The essential information returned by our identification technique for every detected stellar complex is its size, $S$, and the portion of the photometric catalogue 
that corresponds to the stellar sources included within its border, i.e., the number of its stellar members. 
The physical dimensions of each stellar complex are defined by the borders enclosed by the corresponding isopleths. Each isopleth is used for the 
construction of the {\sl convex hull} of the structure. The convex hull of each complex is primarily used for determination of stellar membership for each complex, and the calculation of its {\em effective} or {\em equivalent} radius, $r_{\rm eff}$, defined as the radius of a circle with the same area as that covered by the convex hull of the structure \citep[e.g.,][]{carpenter00, romanzuniga08}. The latter is a measure of the size of each stellar complex as $S = 2\cdot r_{\rm eff}$. 

The convex hull is secondarily used for the determination of the best-fitting ellipse that represents the morphology of 
the complex (Fig.\,\ref{fig:contour_map} right panel). The major and minor semi-axes, $a$ and $b$, of the best-fitting ellipse provide 
the ellipticity or {\em flattening} for each complex, 
\begin{equation}
\displaystyle \varepsilon = \frac{a-b}{a},
\end{equation} 
which is a measure of its elongation with $\varepsilon \in$\,[0,1]. For a circular structure ($a=b$) it has the value $\varepsilon =$\,0. The ellipses determined in terms of convex hull fitting for each structure are plotted in Fig.\,\ref{fig:contour_map} (right panel) to visualize the identified stellar complexes.

\subsubsection{Structural (second generation) parameters}\label{s:param_determ}

\paragraph*{Stellar Mass.}\label{s:advanced_params}
Considering that NGC\,1566 is located at a substantial distance, the sample of stellar members in each complex suffers from incompleteness due to detector sensitivity limitations and photometric confusion. As a consequence the calculation of the total stellar mass of each structure directly from its limited numbers of observed stellar sources will suffer from these constraints. On the other hand, the total sample of 
observed stellar sources provides a rich inventory, which is sufficient for the construction of the complete young stellar mass function (MF) of NGC\,1566, down to the completeness limit of $\sim$\,20\,M{\solar}. We present the construction of this MF in Appendix\,\ref{s:fieldmf}, where also both the MF of the total stellar population in the stellar complexes, and that of all sources outside the complexes are constructed. The corresponding total stellar mass of each of these samples is estimated through the extrapolation of the corresponding MFs (the derived stellar masses are given in Table\,\ref{tab:mfslopes}). The evaluated total stellar mass provides a measure of how much {\sl actual} mass corresponds to each {\sl observed} stellar source (Appendix\,\ref{s:app_smass}). Based on the total stellar mass derived from the MF extrapolation of the observed young stellar sample in NGC\,1566, this mass amounts to $\sim$\,300\,M{\solar} per observed stellar source. We determine the total stellar mass, $M_{\star}$, of each complex by multiplying this mass with the number of its detected stellar members. The derived total stellar masses of the complexes correlate well with their observed UV brightness (Fig.\,\ref{f:masslight}). This correlation expresses  
essentially the mass--luminosity relation derived from the evolutionary models.

\begin{figure*}
\centering
\includegraphics[width=\textwidth]{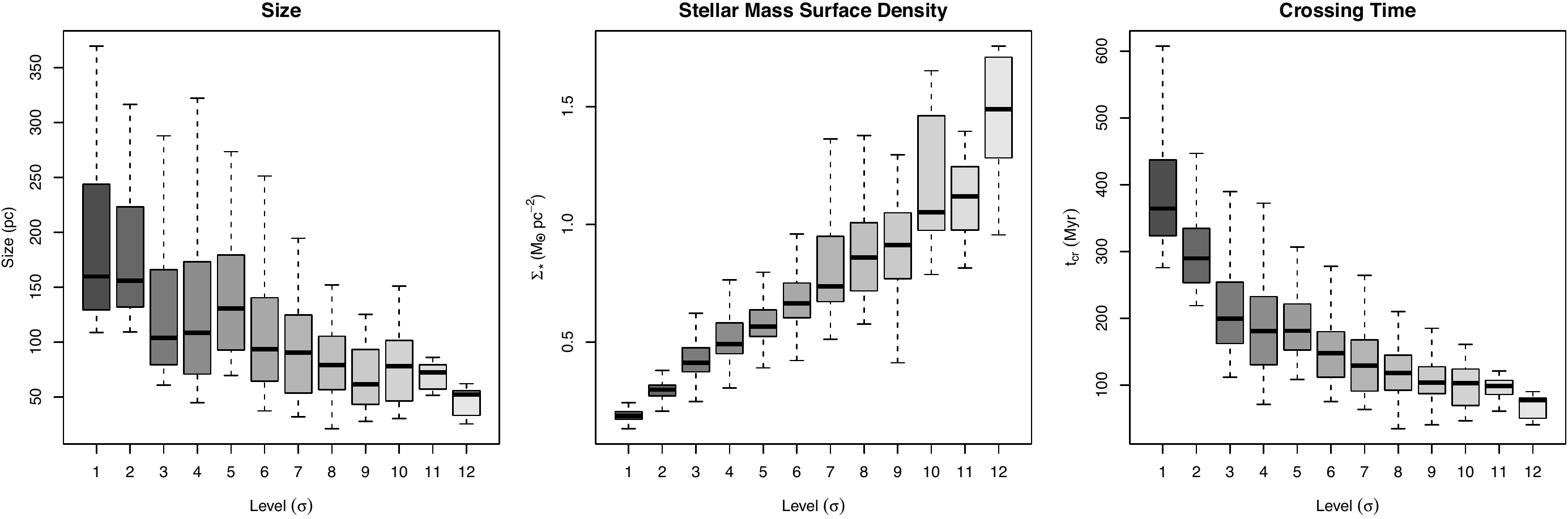} 
\caption{Box plots for three of the parameters derived for the stellar structures detected at different significance density levels. 
In each plot  the median (i.e.,  the 50\% points in cumulative distribution), indicated by the bar in the middle and the interquartile range (i.e, the 
range between the 25\% and 75\% observation), indicated by the box, are compared graphically for each parameter in the various 
sub-samples. These plots visualize the dependence of the structural parameters of the stellar complexes on the detection level for three of the parameters shown in Table\,\ref{tab:structdemo}, namely the sizes, the stellar mass surface densities, and  the crossing times of the systems. \label{fig:boxplots}}
\end{figure*}

\paragraph*{Stellar mass surface density.}

The stellar mass surface density, $\Sigma_{\star}$, of each complex is calculated from its stellar mass and size:
\begin{equation}
\displaystyle \Sigma_{\star} = \frac{M_{\star}}{\pi r_{\rm eff}^2}.
\end{equation}
For this calculation we use the area $\pi r_{\rm eff}^2$, which is identical to the surface covered by the surrounding isopleth of each system 
by definition of $r_{\rm eff}$ (see Sect.\,\ref{s:basic_params}). 

\paragraph*{Dynamical time-scales and velocity dispersion.}

The dynamical status of each stellar complex can be assessed by the  {\sl crossing} 
and the {\sl two-body relaxation}  time-scales \citep[e.g.,][see also \citealt{BinneyTremaine2008, Kroupa2008}]{Spitzer1987}, which are given as:
\begin{eqnarray}
\begin{array}{l}
\displaystyle t_{\rm cr} \equiv \frac{2 r_{\rm eff}}{\sigma_\upsilon}\\
{\rm and}\\
\displaystyle t_{\rm relax} = 0.1 \frac{N_{\star}}{\ln{N_{\star}}} t_{\rm cr}
\end{array}
\label{eq:tcr}
\end{eqnarray}
respectively. The velocity dispersion of the stars in the system, $\sigma_\upsilon$, is estimated from the viral theorem, 
assuming that the systems have come into dynamical equilibrium under gravity:
\begin{equation} 
\displaystyle \sigma_\upsilon \simeq \sqrt{ \frac{{\rm G}M_{\star}}{r_{\rm eff}}},
\label{eq:vdisp}
\end{equation} 
where the gravitational constant has the value ${\rm G} \simeq 4.302 \times 10^{-3}$\,pc\,M$_{\odot}^{-1}$\,(km/s)$^{2}$. 
Our calculations derive a wide range of values for $\sigma_\upsilon$ between 0.3 and 1.7\,km\,s$^{-1}$. 
Considering that stellar complexes are normally not bound, these $\sigma_\upsilon$ measurements are the lower limits 
of the true velocity dispersions of the structures (at least for those at the lowest levels). The estimation of $\sigma_\upsilon$ allows the evaluation of 
$t_{\rm cr}$ and $t_{\rm relax}$ according to Eqs.\,(\ref{eq:tcr}). Box plots for three of the derived parameters, namely size, stellar surface density, and crossing time, are 
shown in Fig.\,\ref{fig:boxplots}. In the following section we present the demographics of the detected complexes, based on their derived parameters. 

\subsection{Parameter demographics}\label{s:param_demo}

The population demographics of all stellar structures revealed at various significance levels, is given in Table\,\ref{tab:structdemo}. 
In the first and second column of the table we give the significance level and the corresponding number of detected structures. The parameters presented in the table include the minimum, average and maximum size (Cols.\,3-5), and the average stellar mass surface density  (Col.\,6) 
of all structures found in each density level. The total stellar mass and total UV magnitudes of all structures in each level are also given (Cols.\,7 and 9 respectively). In Table\,\ref{tab:structdemo} we also provide the fraction of stellar mass $f_{\star}$ located at 
each significance level with respect to the total mass of the whole stellar sample (Col.\,8), and the corresponding stellar UV flux fraction $f_{\rm UV}$ relative to the total observed UV flux per detection level (Col.\,10). Finally the average crossing times and the derived velocity dispersions (assuming that the systems are in virial equilibrium; see  Appendix\,\ref{s:advanced_params}) per significance level are given in Cols.\,11 and 12.


In general, Table\,\ref{tab:structdemo} exhibits a dependence of all basic parameters of the structures on the detection level\footnote{Among all parameters,  ellipticity (see Sect.\,\ref{s:basic_params}), not shown in Table\,\ref{tab:structdemo}, and velocity dispersion appear independent of the detection level.}. The sizes of the complexes cover a wide range from $\sim$\,30\,pc of the smallest 12$\sigma$ stellar aggregate up to scales of over 1\,kpc for the largest 1$\sigma$ super-complexes in the sample. The sizes of the structures decrease while their densities increase with increasing detection level, i.e., those found at higher significance levels become smaller and more dense. 
This trend is visualized with the box plots of these parameters, shown in Fig.\,\ref{fig:boxplots} (left and middle panels). In these box plots the 
parameters of complexes detected at the various density levels are represented by a box of length equal to the interquartile range of the measurements (between the first and third quartiles) and the median of the data\footnote{In the box plots of the figure the upper whisker for parameter $x$ is located at $\min{(\max{(x)}, Q_3 + 1.5  IQR)}$ and the lower whisker at $\max{(\min{(x)}, Q_1 - 1.5 IQR)}$, where $IQR = Q_3 - Q_1$, is the  interquartile range, i.e., the box length.}.  Both the total stellar mass and total UV brightness show a systematic decrease with the detection significance level, with larger and sparser stellar structures hosting higher stellar numbers, masses and UV brightness. This agreement in the trends of these parameters can be directly explained by their strong correlation, as derived from the data of Table\,\ref{tab:structdemo} (see also Fig.\,\ref{f:masslight}).

A systematic dependency on detection level is also obvious for the fraction $f_{\star}$ of stellar mass included in every density level over the total observed mass of the blue stars. This fraction changes from $\sim$\,50\,per\,cent within the 1$\sigma$ structures to $\sim$\,3\textperthousand\, within structures found at the highest density level. Since the 1$\sigma$ isopleths by definition incorporate all the stars that are members of any of the detected stellar complexes (found at various detection levels), and since all of our complexes happen to be in spiral arms, the 50\,per\,cent stellar mass  fraction corresponds to the stellar mass formed along the arms of NGC\,1566 during the last $\sim$\,10 to 20\,Myr. We elaborate more on the significance of this fraction in the discussion of Sect.\,\ref{s:discussion}, where we also discuss the remaining fraction of the total stellar mass being in the ``field'' outside the stellar complexes, still distributed along the arm features of the galaxy.

The dependence of the fraction $f_{\rm UV}$, i.e., the UV emission coming from the structures, on significance level 
is similar to that of the stellar mass fraction $f_{\star}$. There is also a systematic scaling between the two fractions with $f_{\rm UV}$ being almost 1.5 times larger than  
the $f_{\star}$ for all detection levels. As shown in Table\,\ref{tab:structdemo}, crossing times also scale (almost linearly) with the detection 
threshold of the structures. Larger stellar complexes, found at the lower density levels, have systematically longer $t_{\rm cr}$ than the smaller
structures found at higher levels. 

In general all measurements show that $t_{\rm cr}$ is much longer than the covered CMD age 
of $\sim$\,20\,Myr, suggesting that overall the youngest stars in the detected stellar complexes are not mixed. Crossing times longer than stellar ages would indicate that the low-density complexes are 
unbound, i.e., not in virial equilibrium \citep[see, e.g.,][for the distinction between bound star clusters from unbound associations based on the ratio of their 
CMD age over their crossing time]{PortegiesZwart2010}. Nevertheless, that is not necessarily true for the unlike case of large complexes that host multiple star formation events, and therefore stars older than 20\,Myr, which are not detected in our blue CMD. The dynamical two-body relaxation time-scales of the structures, estimated as described in 
Sect.\,\ref{s:advanced_params} (not shown in Table\,\ref{tab:structdemo}), are found to be remarkably high, which indicates that these systems will practically never relax though two-body encounters.



\begin{figure}
\centerline{\includegraphics[clip=true,width=0.475\textwidth]{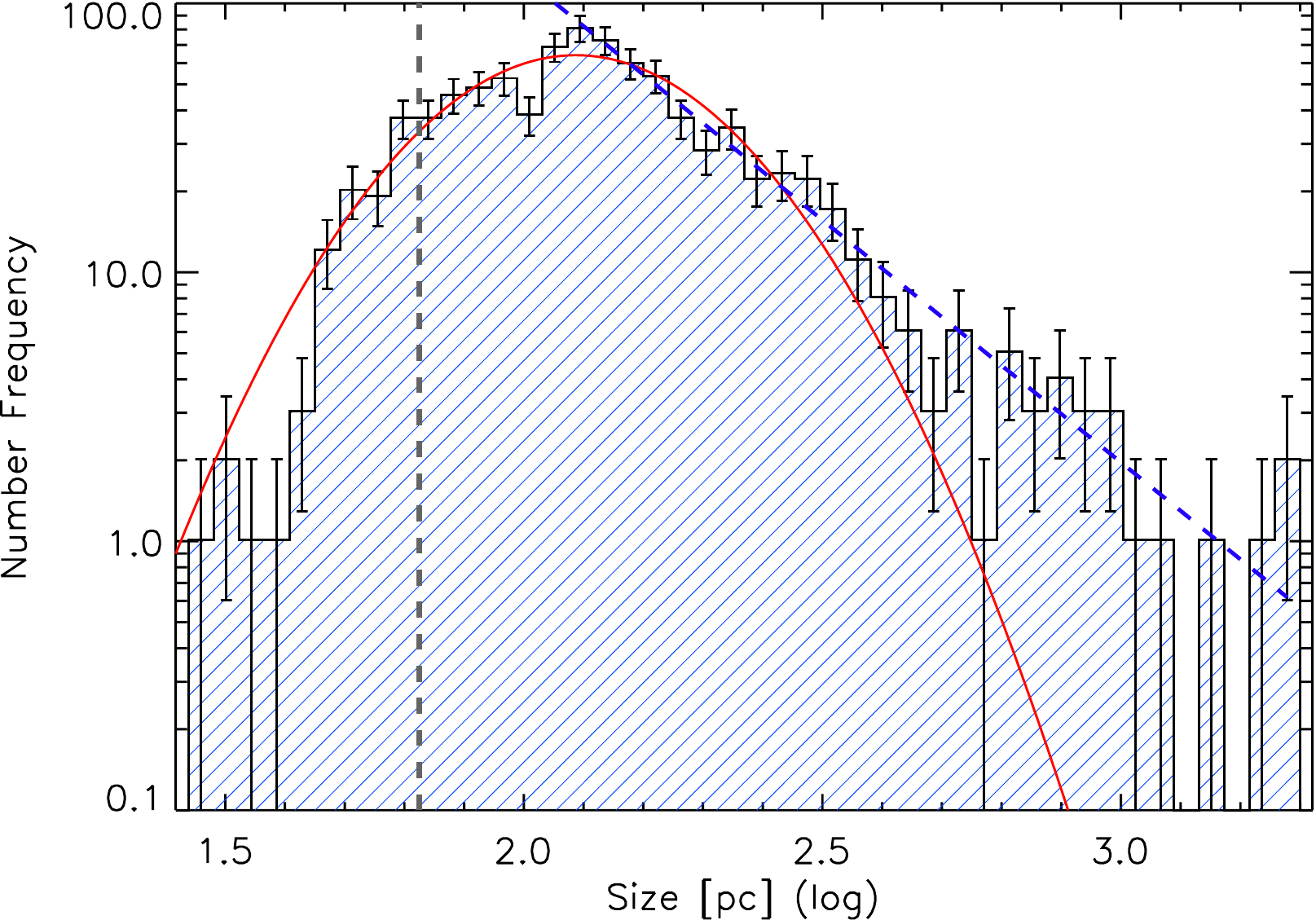}}
\caption{Histogram of the size distribution of all detected young stellar structures in NGC\,1566, 
with a logarithmic bin size of $\sim$\,0.1. Sizes are defined as $2 \cdot r_{\rm eff}$ and are 
given in pc. The histogram peaks at a size of $\sim$\,125$\pm13$\,pc. The best-fit Gaussian 
(red line) peaks at an average size of $\sim$\,122\,pc. The smoothing radius used for the 
structures identification of 67\,pc is indicated with the vertical dashed line. This plot shows 
that the right hand tail of the size distribution follows a power law, 
indicated by the blue dashed line.
\label{f:sizedist}}
\end{figure}

\subsection{Size Distribution}\label{s:size} 

The size distribution of all 890 detected structures is shown in Fig.\,\ref{f:sizedist}. This distribution is constructed by binning all 
systems according to the logarithm of their dimensions, derived from the effective radii of the structures. The dimensions of the  
systems are clustered around an average of $\simeq$\,122\,pc, derived from the functional fit of the histogram to a log-normal distribution 
(drawn with a red line in the figure) with the form:
\begin{equation}
f(S) = \frac{h}{S \sigma \sqrt{2\pi}} \exp\left[-\left(\frac{(\ln S - \mu)^2}{2\sigma^2}\right)\right], 
\label{eq:gaussfit}
\end{equation}                                                       
where $S$ is the size (in pc), and $\mu$, and $\sigma$ are the mean and standard deviation of the natural logarithm of the variable.
The height of the distribution is given by $h$. The derived mean stellar complex size of more than 100\,pc corresponds to that of the largest giant molecular clouds in the Milky Way \citep[e.g.,][]{cox2000, tielens05, HeyerDame2015}. Stellar complexes with this or larger sizes compare more to {\sl cloud complexes} or {\sl conglomerates of clouds} \citep[e.g.,]{Grabelsky1987}, large structures of molecular clouds with extended atomic gas envelopes \citep[H{\sc i} superclouds; e.g.,][]{EE1987}. The sizes of complexes on the left wing of the distribution of Fig.\,\ref{f:sizedist} are comparable to those of typical stellar associations and aggregates in Local Group galaxies \citep[e.g.,][]{efremov87, ivanov96}. 


\begin{figure}
\centerline{\includegraphics[clip=true,width=0.475\textwidth]{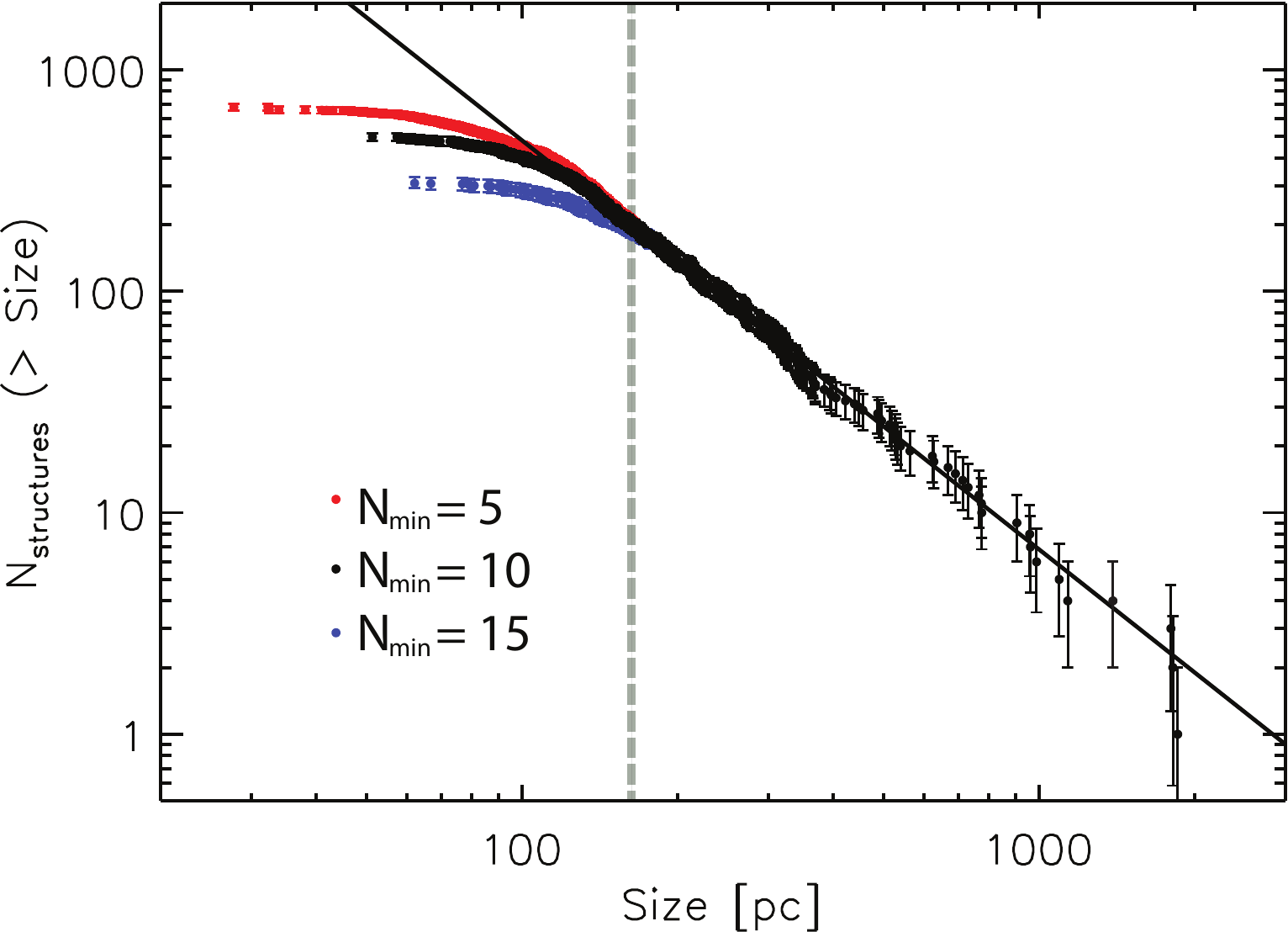}}
\caption{Cumulative size distribution of the detected stellar complexes for three $N_{\rm min}$ limits used in our identification technique. The size beyond which all three distributions have the same shape ($\sim$\,165\,pc, indicated with the vertical dash line) specifies the completeness limit of our detection. Beyond this size the distributions are in practice identical and follow a clear power-law of the form $N \propto S^{-\beta}$ with an exponent $\beta \sim$\,1.8\,$\pm$\,0.1 (shown with the continuous black line). To avoid confusion, only the data points for $N_{\rm min} = 10$ are shown beyond the completeness limit. Error bars correspond to counting errors. 
\label{f:cumszdist}}
\end{figure}

The Gaussian fit in Fig.\,\ref{f:sizedist} demonstrates that the size distribution of the detected systems 
is not entirely log-normal. At the right-hand part of the distribution there is an overabundance of large structures 
with respect to the best-fitting Gaussian. We illustrate this effect by fitting a power-law for sizes larger than the mean. 
Moreover, the left-hand part of the distribution is certainly affected by our detection limit.
Indeed, the size distribution of the detected structures may be affected by incompleteness in our identification. There are two parameters considered in our technique, which affect our detection completeness: The KDE kernel applied for the construction of the  stellar surface density map and the minimum number of members in defining a structure (we used $N_{\rm min} = 5$).  Our analysis 
of stellar complexes in NGC\,6503 showed that the peak in the size distribution does depend on the resolution of the detection 
technique, i.e., the KDE kernel size \citep{Gouliermis2015}, but this dependence accounts for no more than $\sim$\,10\,per\,cent differences.
Moreover, the resolution we use here for NGC\,1566 is the highest allowed in order to avoid significant noise levels in the
stellar density maps, and therefore the derived average size is the smallest that can be resolved at the distance of NGC\,1566. 

In order to quantify the effect of the choice of $N_{\rm min}$ to our completeness and to the shape of the size distribution, we constructed this distribution for different 
$N_{\rm min}$ values. We found that the size at the peak of the distribution does not change with higher $N_{\rm min}$, but the height of the distribution lowers.  
More importantly, the power-law tail of the distribution was found to remain prominent also for higher $N_{\rm min}$ limits, while becoming somewhat flatter. This is further demonstrated by the cumulative size distribution of the detected complexes, shown in Fig.\,\ref{f:cumszdist} for three $N_{\rm min}$ limits (5, 10, and 15 stars). In a recent study of hierarchical star formation in the 30 Doradus complex \cite{Sun2017ApJ} show that the cumulative size distribution of the detected stellar groups does change with $N_{\rm min}$ at small scales, but the power-law part at larger sizes remains unchanged. Indeed, in Fig.\,\ref{f:cumszdist}  it is shown that while the left-hand (small scales) part of the cumulative size distribution flattens with higher $N_{\rm min}$, the power-law tail at the right-hand (large scales) part remains unaltered. The average best-fitting power-law to all distributions has an exponent $\beta =$\,1.8\,$\pm$\,0.1. The power-law tails seen in both the differential and cumulative size distributions clearly suggest a hierarchical mechanism in determining the sizes of the stellar complexes. We test this hypothesis assuming a ``hierarchical fragmentation'' toy-model, described in the next section.

\begin{figure}
\centerline{\includegraphics[clip=true,width=0.475\textwidth]{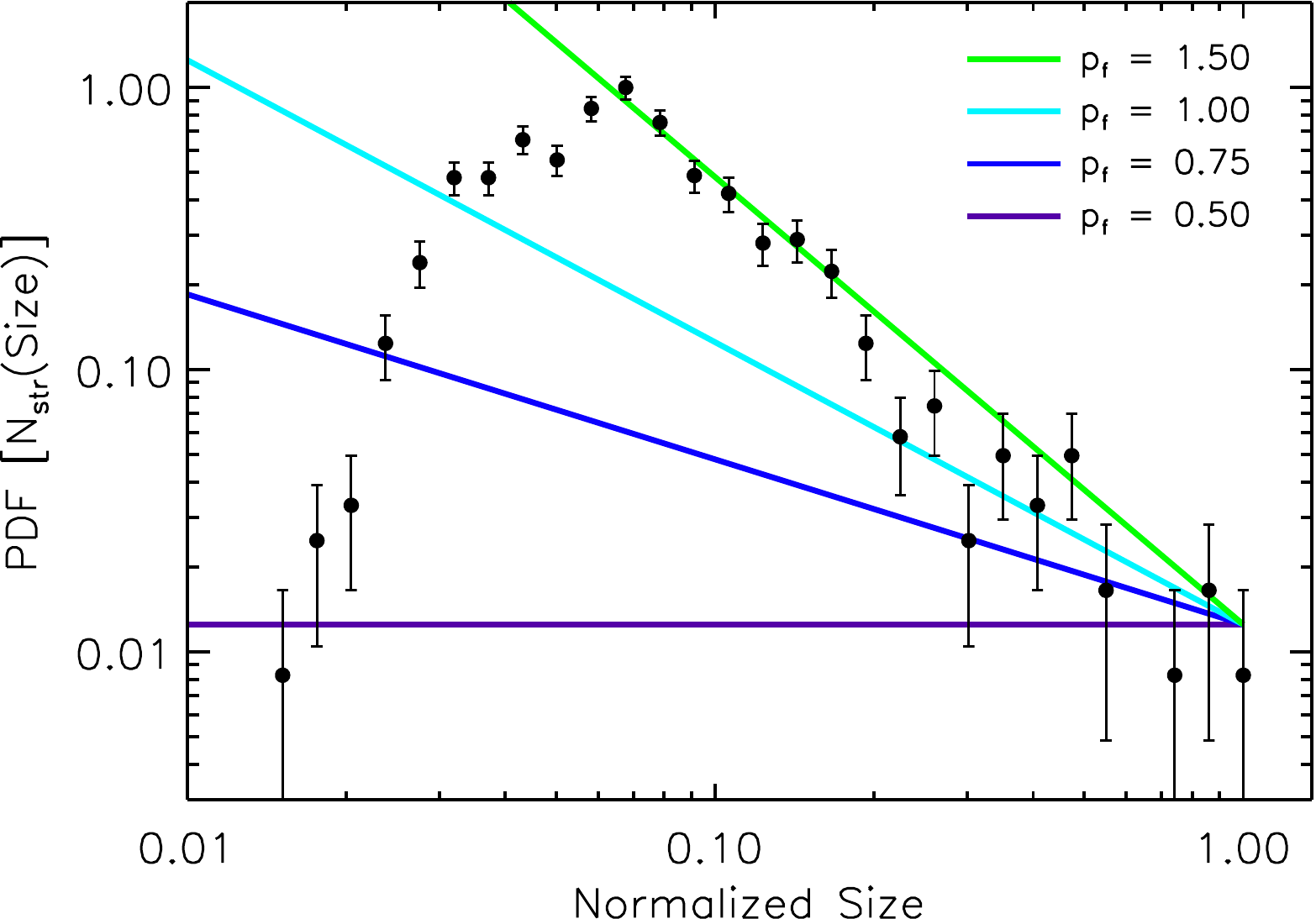}}
\caption{Size probability distribution function of the total sample of star-forming complexes detected in NGC\,1566,
i.e. for the structures identified at density levels $\geq$\,1$\sigma$. The predictions of a {\sl fragmentation and enrichment}  
model for the power-law tail of the size PDF overlaid with different colours for different fragmentation probabilities (see Sect.\,\ref{s:toymodel}). 
\label{f:pdfrad}}
\end{figure}

\subsubsection{A model for the power-law tail of the size PDF}\label{s:toymodel}

We build a naive model to explain the origin of the right-hand (toward large sizes) power-law tail observed in the size PDF of the total sample of 
stellar complexes in NGC\,1566, illustrated in Fig.\,\ref{f:pdfrad}. 
Our analysis has shown thus far the hierarchical connection between larger and smaller young stellar structures.  
But how is this organized? 
Let us assume an original sample of large stellar complexes, some of which ``fragment'' into smaller (and denser) 
sub-structures. The fraction of stellar complexes that ``break" into smaller systems is defined by a {\sl fragmentation probability} $p_{\rm f}$, 
i.e., the probability that any given complex will eventually fragment. The number of the sub-structures that will be produced 
in every fragmented complex is given by $n_{\rm s}$. Fragmentation is treated in the simplest way, as a multiplicative process 
in which the sizes of substructures are fractions of the size of the original \citep[e.g.,][]{Sornette2004}. The size of each of the substructures in 
every fragmented complex is thus defined as a fraction of the original complex's size $f_{\rm s}  < 1$. If we assume that the total size of 
the sub-structures cannot exceed 
that of the parental structure, the latter is expressed as $f_{\rm s} = 1/n_{\rm s}$. 

For an original sample of $N_0$ complexes of size $S_0$, the number of  substructures and their corresponding 
sizes for every  $i^{\rm th}$  ``generation'' of sub-structures derived from the fragmentation of the previous generation is given as:
\begin{eqnarray}
\begin{array}{rl}
\displaystyle {\rm Number\, of\, Structures: }& N_i = (p_{\rm f} n_{\rm s})^i \cdot N_0 \\
\\
\displaystyle {\rm Size\, of\, Structures: } & S_i = f_{\rm s}^i \cdot S_0 \equiv {n_{\rm s}}^{-i} \cdot S_0,
\end{array}
\label{eq:series}
\end{eqnarray}
where $i \in$\,[0, 1, 2, ...]. This kind of simple fragmentation models produce power-law shaped size PDFs. 
Indeed, based on fragmentation phenomena on Earth, most of the size distributions of fragments (not conditioned by 
a given generation rank) display power-law behaviour $P(x) \propto x^{-\tau}$ with exponents $\tau$ between 1.9 and 
2.7 \citep{Turcotte1986}. Four examples of the power-laws produced by the model for various values of $p_{\rm f}$  
and a fixed value for $f_{\rm s} = 0.5$ (i.e., $n_{\rm s} =$\,2) are shown in Fig.\,\ref{f:pdfrad}, overlaid on the observed size 
PDF. A fragmentation probability of 0.5, 0.75, 1.00 and 1.50 is considered in the examples. This figure shows that a simple 
(naive) fragmentation model may explain the power-law tail in the size PDF of stellar complexes in NGC\,1566. 

From the modelled power-laws shown in Fig.\,\ref{f:pdfrad}, those corresponding to $p_{\rm f}=$\,0.75 and 1 
seem to fit well the large-scale end of the distribution, implying a fragmentation probability for the large structures that varies between these 
values. On the other hand, the small-scale part of the tail (up to the peak of the distribution) is better represented by 
the unrealistic\footnote{The value  $p_{\rm f}=$\,1.5 is not realistic, since there cannot be more objects fragmented than those available in the parental sample.} fragmentation probability of $p_{\rm f}=$\,1.5. While this value being $p_{\rm f} >$\,1 is impossible, it simulates the {\sl enrichment} of every generation with {\sl new objects} appearing in addition to those produced by fragmentation of objects in the previous generation. 
Specifically, the value of $p_{\rm f}=$\,1.5 resembles the case where 100 per cent of the objects in the parental generation 
will fragment, while a number of new objects equal to the number of the original parental sample ( $2 \times 50$\,per\,cent) will
be added to the new generation by ``external'' mechanisms. The latter correspond to formation events of new (small) stellar complexes, 
driven by, e.g., turbulence or other global processes.  

The ``enrichment'' process described above resembles that proposed by \cite{Yule1925} to explain the distribution of the number of species in a genus, family or other taxonomic group \citep{WillisYule1922}. Processes like that, where new objects appear in between the appearance of one generation and the next are known as ``rich-get-richer'' mechanisms \citep[e.g.,][]{Simon1955}. In their derived distributions, which appear to follow power laws quite closely, the probability of a generation gaining a new member is proportional to the number already there \citep[see, e.g.,][]{NetworksCrowdsMarkets2010}. The Yule process, along with systems displaying self-organized criticality\footnote{In systems with self-organized criticality a scale-factor of the system diverges, because either the system is tuned to a critical point in its parameter space or it ``automatically'' drives itself to that point. The divergence leaves the system with no appropriate scale factor to set 
the size of the measured quantity, which then follows a power law.} are considered to be the most important physical mechanisms for the occurrence of power laws \citep[see][for a review]{Newman2005}.

\begin{figure}
\centerline{\includegraphics[clip=true,width=0.475\textwidth]{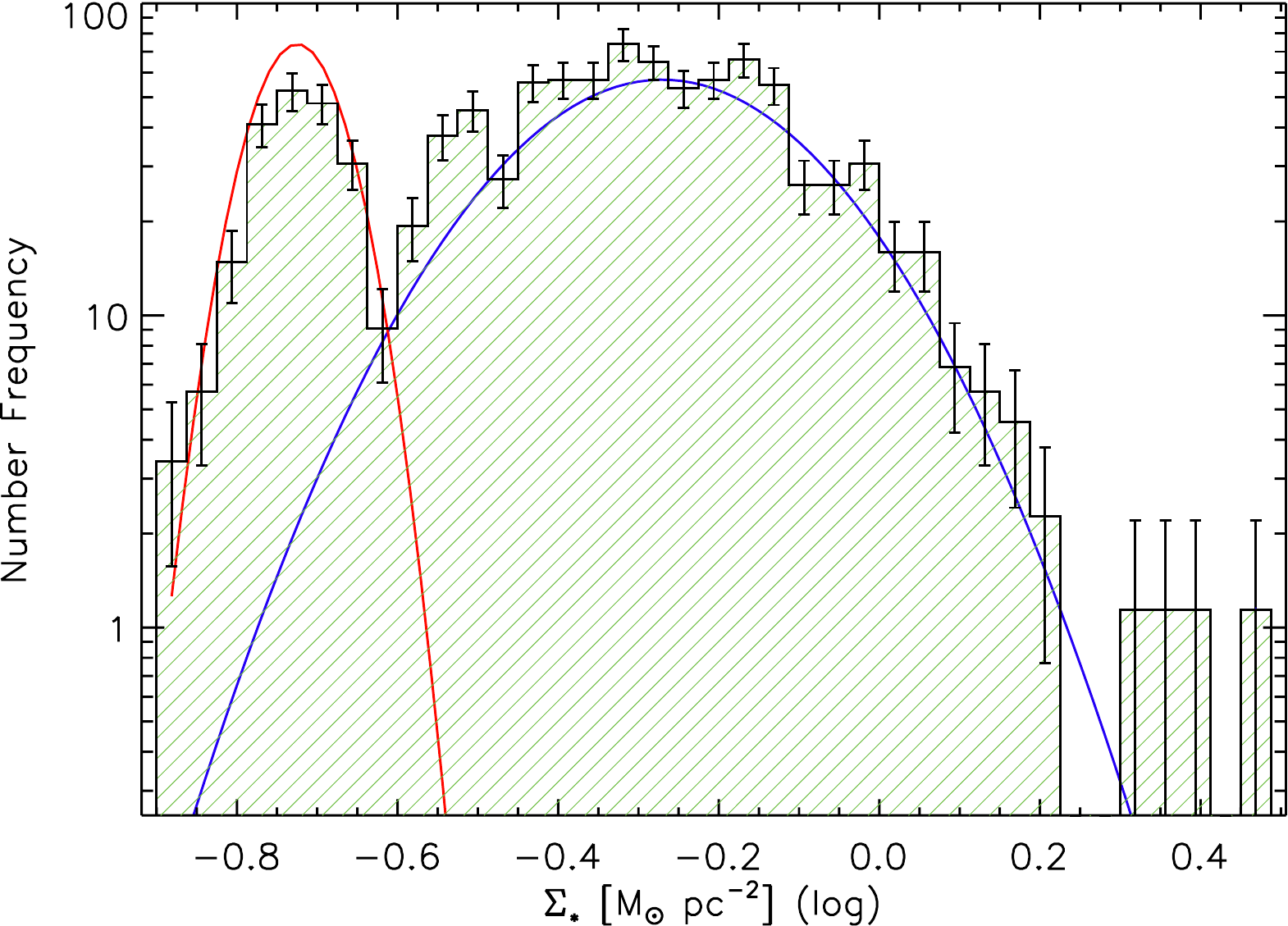}}
\caption{Histogram of the surface stellar mass density distribution of all detected young 
stellar structures in NGC\,1566, with a logarithmic bin size of $\sim$\,0.1. This distribution has a bimodal shape, 
with the first narrow mode being produced by the 1$\sigma$ structures (red line), while 
the second by the $\geq$\,2$\sigma$ complexes (blue line). Both modes are represented 
by log-normal distributions, depicted in different colours by their best-fitting Gaussians.
\label{f:surfdensdist}}
\end{figure}

In our naive hierarchical fragmentation and enrichment model we assume both the fragmentation probability and the size fraction, as well as
the fraction of newly added members in every generation to be constant, i.e., they remain the same for all generations. However, it is quite  
possible that this convention may not apply in real stellar complexes, since one may expect a dependence of these parameters on the typical 
characteristics of structures in each generation. More sophisticated models possibly recreate sensible samples of hierarchically formed stellar structures \citep[see, e.g.,][for an analytic framework of fragmentation in turbulent, self-gravitating media\footnote{In particular, Sect.\,11 and Fig.\,12 in \cite{hopkins2013} describe the `fragmentation trees' of  collapsing molecular clouds.}]{hopkins2013}. The naive model discussed here provides a reasonable simple scenario for the power-law tail in the size PDF of young stellar complexes observed in NGC\,1566.

\begin{figure*}
\centerline{\includegraphics[clip=true,width=0.875\textwidth]{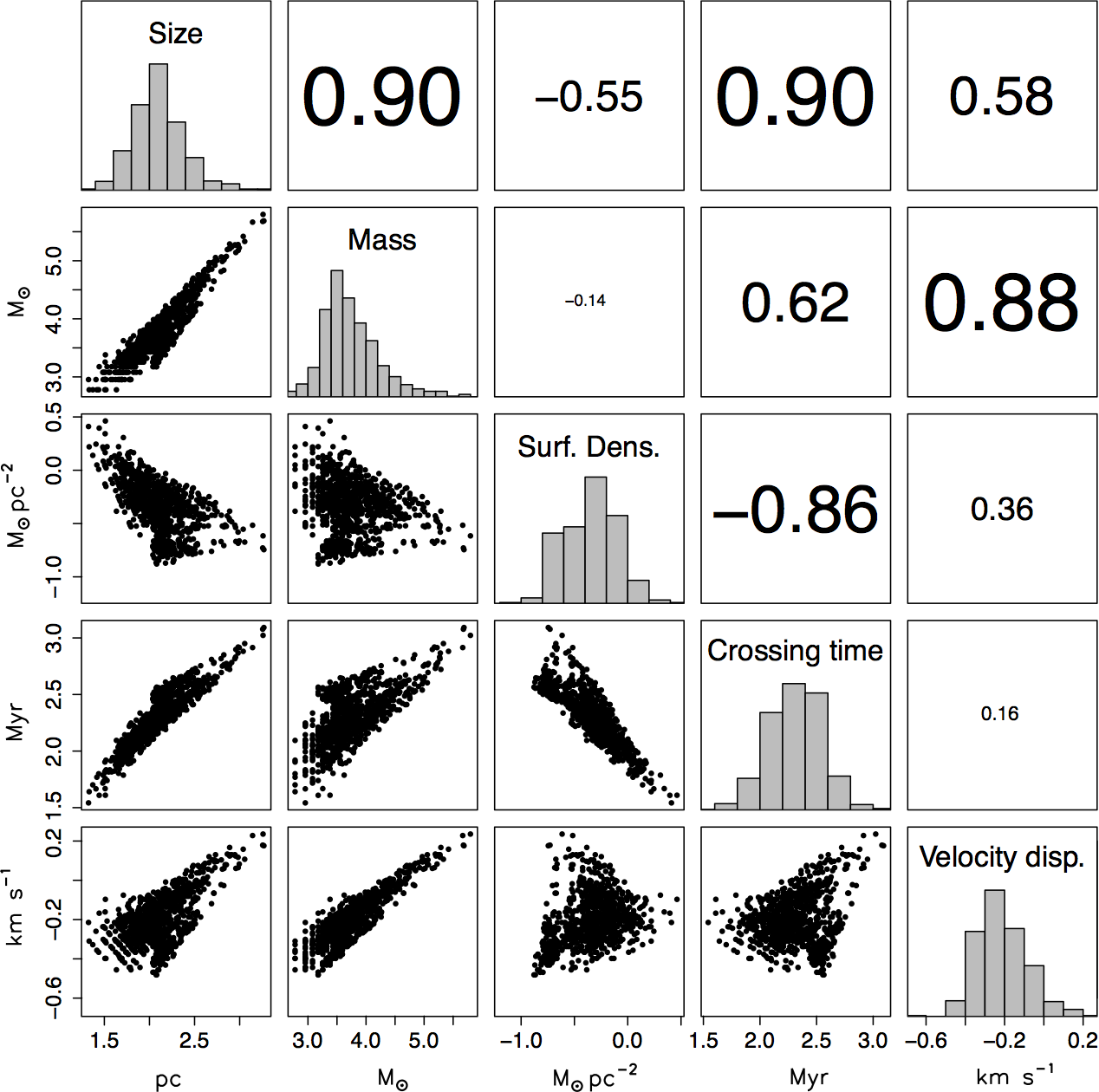}}
\caption{Overview of the correlations between the structural parameters (in logarithmic base 10 scale) of the 
stellar complexes in NGC\,1566. Scatter plots for the entire sample of structures, as found in all detection levels, are shown. 
The bivariate scatter plots are shown below the diagonal, and the histograms of the parameters are on the diagonal. 
The corresponding Pearson correlation coefficients are given for every parameters pair above the diagonal. All
values are given in logarithmic scale base 10. 
\label{f:paramcorrel}}
\end{figure*}

\subsection{Surface Stellar Mass Density Distribution}\label{s:surf_dens_distr} 

The stellar mass surface density distribution of the identified structures behaves differently than 
their size distribution, as demonstrated in Fig.\,\ref{f:surfdensdist}. This distribution shows a bimodal shape, 
which is invariable with the choice of bin size. We verified that the first narrow mode is entirely 
produced by the 1$\sigma$ structures, while the structures in all remaining density levels produce the second mode 
in the distribution\footnote{We treat this distribution as the mixture of two unimodal distributions, each well fitted by a 
log-normal distribution, depicted in Fig.\,\ref{f:surfdensdist} with different colours.}. This behaviour of the density distribution 
with the 1$\sigma$ structures in a separate mode and all remaining structures being clustered under a common  
log-normal distribution, suggests a clear distinction between the 1$\sigma$ and $\geq$\,2$\sigma$ complexes  
in terms of stellar density scale. We verified that this bimodal behaviour, as well as the log-normal shape which is discussed below, 
remain unchanged for different $N_{\rm min}$ limits.

Both density distributions are well represented by log-normal functional forms similar to that of Eq.\,\ref{eq:gaussfit}  
(with $S$ being replaced by $\Sigma_{\star}$). The corresponding best-fitting Gaussians (plotted  in 
Fig.\,\ref{f:surfdensdist} with a blue and a red line) peak at $\sim$\,0.2\,M{\solar}\,pc$^{-2}$ for the 1$\sigma$ complexes and at $\sim$\,0.5\,M{\solar}\,pc$^{-2}$
for the remaining structures. Hydrodynamical simulations have shown 
that the column density PDF for supersonic non-gravitating turbulent gas in an isothermal environment has a log-normal form 
\citep[e.g.,][]{Vazquez1994, Padoan1997, Federrath2010, Konstandin2012}. With self-gravity becoming important due to star formation, 
the number of dense regions increases and this introduces a power-law tail on the high-density side of the PDF \citep[e.g.,][]{Klessen2000, 
VazquezSemadeni2008, Collins2012, Girichidis2014}. These predictions are observationally verified for giant molecular clouds (GMCs) in the Milky Way \citep[e.g.,][]{Lombardi2010, Schneider2012}. Log-normal distribution of stellar surface density has also been reported by \cite{Bressert2010} for young stellar objects in the solar neighbourhood. Considering extragalactic environments, the CO emission PDFs for the inner disk of M\,51, M\,33 and the Large Magellanic Cloud are also found to be represented by log-normal functions \citep[][]{Hughes2013}. Considering these findings, the log-normal shape of our stellar surface density PDF suggests that the observed stellar density distribution of the detected structures may be the product of turbulence, in accordance to column density investigations of the interstellar medium.

If the stellar surface density of the star-forming complexes in NGC\,1566 is indeed linked to the gas density of their GMCs,
then the log-normal shape of its PDF may be explained as being inherited by the molecular gas properties. A theoretical framework to explain the log-normal shape and the appearance of a power-law tail at high densities in the PDFs for turbulent self-gravitating clouds is developed by \citet{Elmegreen2011}, using convolution PDFs that depend on the maximum to minimum (i.e., the core-to-edge) average cloud density ratio.  
According to this model, if there is a critical column density for star formation, then the fraction of the local mass exceeding this threshold becomes higher near the cloud centre and bound structures form there due to high efficiency. The fact that the surface stellar mass density PDF of the NGC\,1566 complexes does not show a significant power-law tail implies that self-gravity effects are not visible in this PDF, since sensitivity and resolution limitations do not allow the detection of the highest-density compact star-forming centres of the complexes. As a consequence, this PDF shows only the outcome of global turbulence-driven effects. 


\subsection{Parameter correlations}\label{s:param_correl}

Correlations between observed parameters are powerful tools in understanding the physical conditions of 
various phenomena in astronomy. 
The characteristics of the identified star-forming complexes  are tightly connected to the properties of the molecular 
gas in the galaxy, its disk dynamics and the star formation process itself (see, e.g., discussion in Sect.\,\ref{s:intro}).
We investigate the structural morphology, and thus the conditions of the formation of the detected structures from the correlations between their 
measured  structural parameters. The basic parameters considered in our analysis are the size of the structures ($2 \cdot r_{\rm eff}$), their 
surface stellar mass density, $\Sigma_{\star}$ (in M{\solar}\,pc$^{-2}$), the total stellar mass, $M_\star$, as well as their 
crossing times $t_{\rm cr}$ (in Myr). All parameters are derived from the observed stars in each structure, as described in Sect.\,\ref{s:params}.

The overview of the scatter plots between these parameters is presented in Fig.\,\ref{f:paramcorrel}. The 
correlations are shown below the diagonal of the plot. Histograms of the considered parameters are 
shown on the diagonal. Above the diagonal we provide the corresponding Pearson correlation coefficients. 
From these scatter plots it is shown that strong correlations (or anticorrelations) exist between the mass and size 
of the systems. Also the crossing time appears to be well-correlated with both the size and the stellar mass 
surface density, and the velocity dispersion to correlate well with mass. Weaker correlations 
exist between crossing time and mass, velocity dispersion and size, and surface stellar density and size.
Surprisingly, no significant correlation can be seen between the surface density and the mass, and between velocity 
dispersion and crossing time.

One of the strongest correlations between the derived parameters, shown in Fig.\,\ref{f:paramcorrel}, is that between 
size and crossing time  (Eqs.\,\ref{eq:tcr} and \ref{eq:vdisp}
in Sect.\,\ref{s:params} for the functional relations between these parameters), which however can 
be explained {\sl by definition}, assuming constant (or slowly varying) velocity dispersion. Among all correlations,
of particular interest to our analysis are those between parameters {\sl not} related to each other by definition.
The most prominent is the mass--size 
relation, which shows a positive dependence between these parameters with more stellar mass being accumulated in the larger structures. We explore 
this trend, which is equivalent to that observed in molecular clouds populations, in the following section. 
Another interesting 
relation is that between the
stellar surface density and size, which is not as strong, but it influences the mass--size relation of the structures. Other relations we take a closer
look at are those between crossing time and surface density, which relate to each other through size, and between velocity dispersion and size  (Sect.\,\ref{s:rho_sz}).

\begin{figure}
\centerline{\includegraphics[clip=true,width=0.475\textwidth]{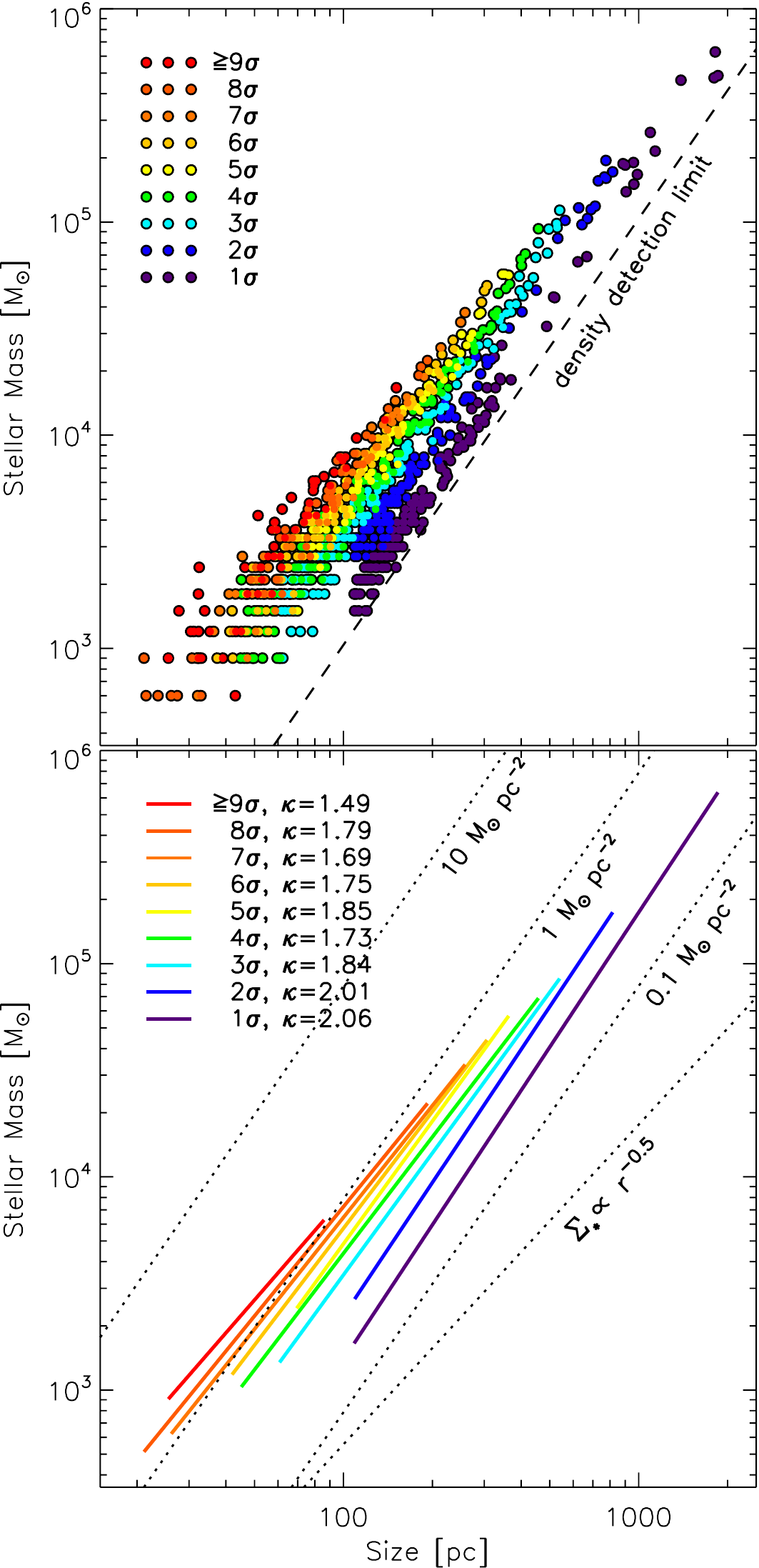}}
\caption{Mass--size relation for the detected stellar structures. {\em Top:} 
Scatter plot for all systems identified at various density levels. Points with different colours represent 
different groups of systems as indicated in the legend. The dashed line gives the average surface 
mass density (which in our analysis is the detection surface density limit). {\em Bottom:} The best 
fitting power-laws of the mass--size relation for systems 
found at various significance levels. Lines with different colours represent different groups of systems 
as indicated in the legend, where the corresponding power-law index, $\kappa$, is also given.
Dotted lines indicate the mass--size relations for uniform stellar mass surface densities (for 0.1, 1, and 
10 M{\solar}\,pc$^{-2}$), and of centrally condensed systems following a power-law 
radial stellar mass surface density profile of the form $\propto r^{-0.5}$.
\label{f:sizenstars}}
\end{figure}

\subsubsection{Stellar-Mass--Size Relation}\label{s:size_mass}

In Fig.\,\ref{f:sizenstars} we show the measured size ($S \equiv 2 \cdot r_{\rm eff}$) versus the stellar mass $M$ within each identified structure. 
The scatter plot of these parameters is shown for all systems in the top panel of the figure. Points corresponding to systems in different detection 
density levels are indicated by different colours. The mass--size relation of all systems, as well as those in individual detection levels, can be 
represented very well by a power-law of the form 
\begin{equation}
M_\star \propto S^{\kappa}. 
\label{fun:msrel}
\end{equation}                  
Fig.\,\ref{f:sizenstars} shows that the mass--size relation of the 
systems depends on their detection limit, with that for the loose (1$\sigma$, 2$\sigma$) structures showing an index corresponding to systems with a 
uniform stellar surface density ($\kappa \simeq$\,2), while systems found at higher density levels show a power-law mass--size relation with a fractional 
index $\kappa <$\,2. 

In general there is no significant scatter in the mass--size relation of the whole sample of stellar complexes. The Pearson correlation coefficient 
of the relation indicates a strong positive relationship (Fig.\,\ref{f:paramcorrel}). The power-law index of the relation for the whole sample is 
$\kappa \simeq$\,1.54. Such indexes are expected for fractal distributions \citep[see, e.g.,][]{elmefalga96}, in 
agreement with results from the application of a different technique on the spiral M\,33 \citep[][]{bastian07}. 
In the bottom panel of Fig.\,\ref{f:sizenstars} the mass--size relations of the systems are shown with solid lines representing the 
corresponding best-fitting power-laws for every group of systems. The corresponding exponents $\kappa$ are indicated in the plot for every group.
They are also given with their Pearson correlation coefficients in Table\,\ref{tab:raramcorr} (Cols. 2 and 3 respectively).
The mass--size relations for structures of constant stellar surface density for three fixed values (i.e., 0.1, 1, and 10 M{\solar}\,pc$^{-2}$) are also shown 
in the plot with dotted lines. That for structures with a radial surface density profile of the form $\displaystyle \Sigma_{\star} 
\propto r^{-0.5}$ is also shown with a dotted line. 

The mass--size relations for systems found at the 1$\sigma$ and 2$\sigma$ density threshold show the power-laws of constant-density systems.
In practice, as shown in the figure, the 1$\sigma$ complexes follow the mass--size relation for a constant density of  $\sim$\,0.2\,M{\solar}\,pc$^{-2}$; those found at 
2$\sigma$ for somewhat higher density. On the other hand structures found at higher density levels show flatter mass--size relations more compatible to that expected for structures with size-dependent stellar surface densities. This trend is consistent to what is found for the mass--size relations of compact young clusters and associations in the Magellanic Clouds 
\citep[see e.g.,][and references therein]{gouliermis03}.  In the simple case, where we assume structures following a power-law surface density dependence on size of the form  $\displaystyle \Sigma_{\star} \propto S^{-\nu}$, the exponent of their mass--size relations connects to their density--size 
exponent as $\kappa = 2 - \nu$. We can thus parametrize the density-size relation with the $\kappa$ exponent of the mass--size relation as we discuss in the following section.

\begin{figure}
\centerline{\includegraphics[clip=true,width=0.475\textwidth]{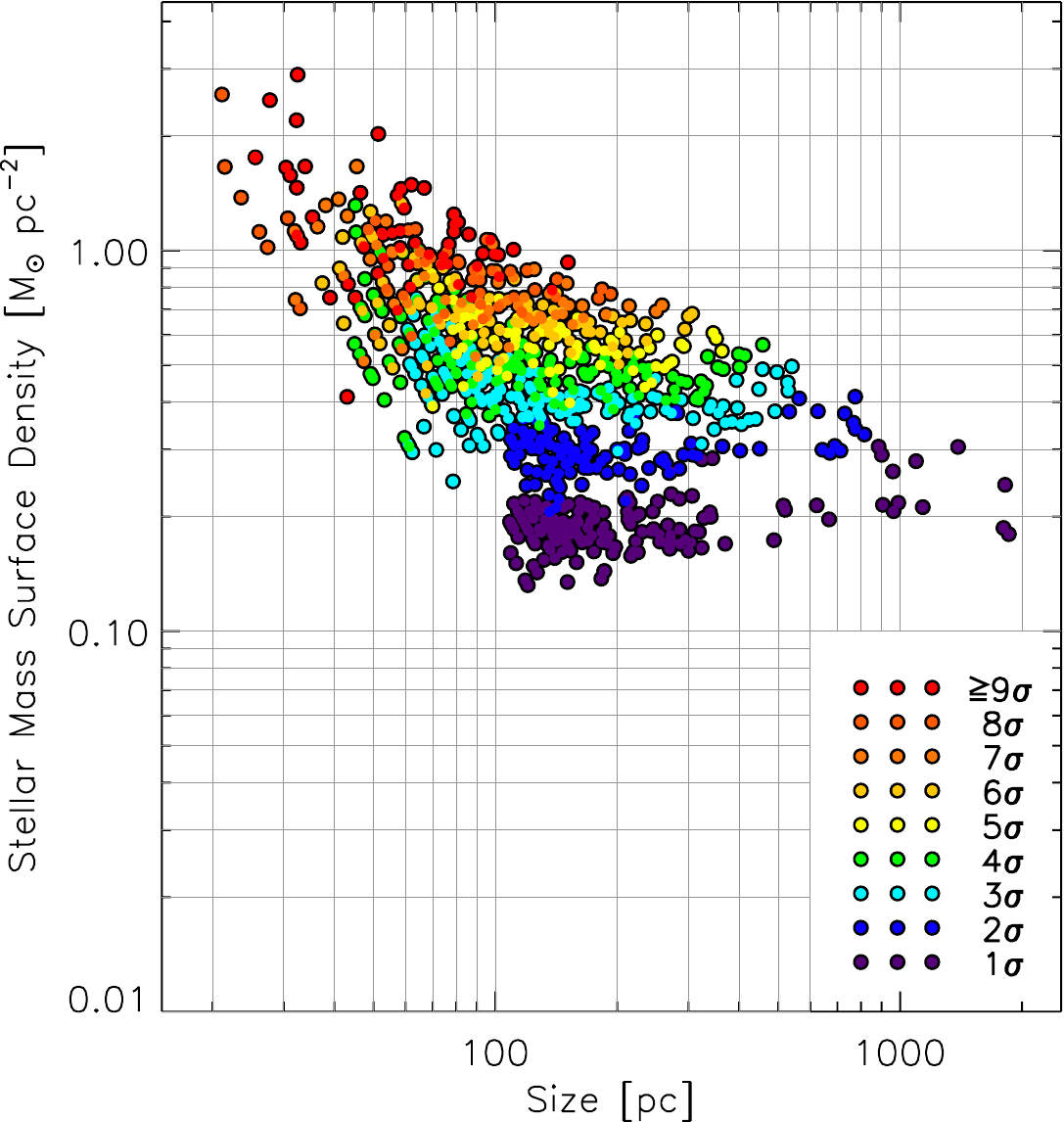}}
\caption{Stellar  mass surface density correlation with the size of the detected stellar structures. 
As in the mass--size relation the strength of the correlation between density and size changes significantly from one group of structures to the other. No significant relationship between these parameters exists for the low-density complexes (blue and purple symbols), for which density is almost constant. In contrast, the correlation of surface density and size becomes progressively stronger for systems detected at higher density levels, with the most compact systems showing the most prominent relationship between these two parameters (red and orange symbols). 
\label{f:sizedenscorr}}
\end{figure}


\begin{table}
\centering
\caption{Power-law exponents ($\kappa$, $\nu$, $\lambda$, $\xi$) and the corresponding Pearson correlation coefficients ($r_{\rm p}$) 
for the correlations with size of four parameters for the stellar complexes in various density levels. The negative sign of exponent 
$\nu$ signifies anticorrelation between surface density and size. All correlations are parametrised by the mass--size relation exponent $\kappa$:
$\nu=\kappa-2$, $\lambda=(\kappa -1)/2$, and $\xi = (3-\kappa)/2$.}
\label{tab:raramcorr}
\begin{tabular*}{\columnwidth}{r  c c r@{\extracolsep{\fill}}c c@{\extracolsep{\fill}}c c@{\extracolsep{\fill}}c}
\hline
\multicolumn{1}{c}{Level} &
\multicolumn{2}{c}{Mass--Size} &
\multicolumn{2}{c}{Density--Size} &
\multicolumn{2}{c}{Velocity--Size} &
\multicolumn{2}{c}{Time--Size}\\ 
\multicolumn{1}{c}{} & 
\multicolumn{2}{c}{$M_\star \propto S^{\kappa}$} & 
\multicolumn{2}{c}{$\Sigma_\star \propto S^{\kappa-2}$} & 
\multicolumn{2}{c}{$\sigma_v \propto S^{\frac{\kappa-1}{2}}$} & 
\multicolumn{2}{c}{$t_{\rm cr} \propto S^{\frac{3-\kappa}{2}}$}\\ 
\multicolumn{1}{c}{} &
\multicolumn{1}{c}{$\kappa$} &
\multicolumn{1}{c}{$r_{\rm p}$}&
\multicolumn{1}{c}{$-\nu$} &
\multicolumn{1}{c}{$r_{\rm p}$}&
\multicolumn{1}{c}{$\lambda$} &
\multicolumn{1}{c}{$r_{\rm p}$}&
\multicolumn{1}{c}{$\xi$} &
\multicolumn{1}{c}{$r_{\rm p}$}\\
\hline 
                1 &  2.06 & 0.99 & $-$0.06 & 0.25 & 0.53& 0.98& 0.47& 0.97\\
               2 &  2.01 & 0.99 & $-$0.01 & 0.26 & 0.51& 0.98& 0.49& 0.98\\
               3 & 1.84 & 0.99 & 0.16 & 0.54 & 0.42& 0.96& 0.58& 0.98\\
               4 & 1.73 & 0.98 & 0.27 & 0.63 & 0.37& 0.91& 0.63& 0.97\\
               5 &  1.85 & 0.99 & 0.15 & 0.46 & 0.42& 0.94& 0.58& 0.97\\
               6 & 1.75 & 0.98 & 0.25 & 0.60 & 0.37& 0.91& 0.63& 0.97\\
               7 &  1.69 & 0.98 & 0.31 & 0.67 & 0.35& 0.90& 0.65& 0.97\\
               8 & 1.79 & 0.98 & 0.21 & 0.49 & 0.40& 0.91& 0.60& 0.96\\
   $\geq$\,9 &  1.49 & 0.95 & 0.51 & 0.72 & 0.24& 0.70& 0.76& 0.95\\
         Total &1.52 & 0.90 & 0.48 & 0.55 & 0.26& 0.58& 0.74& 0.90\\
\hline
\end{tabular*}
\end{table}

\begin{figure*}
\centerline{\includegraphics[clip=true,width=0.75\textwidth]{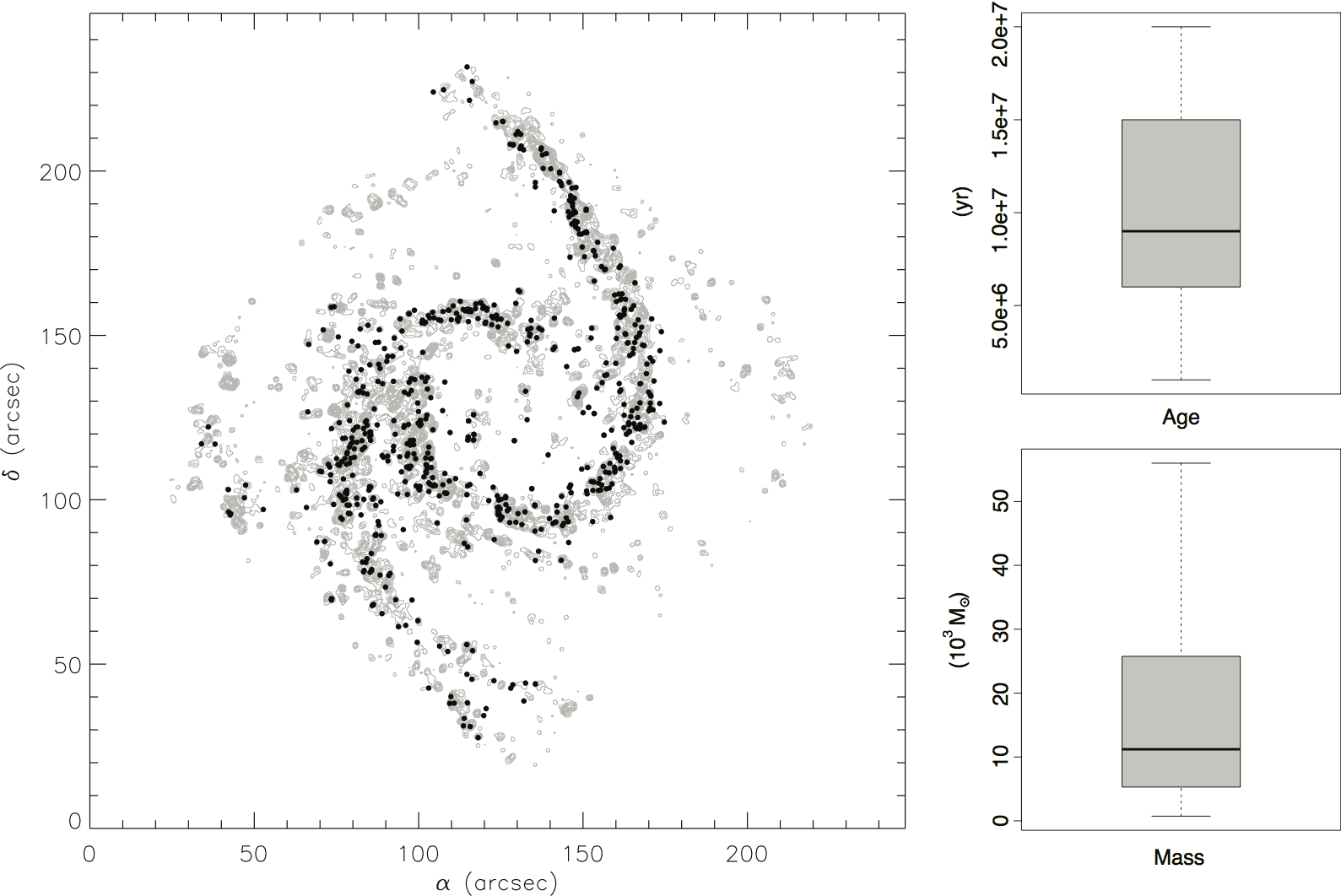}}
\caption{{\em Left panel.} Map of the 625 known young star clusters associated with the stellar complexes and spiral arms of NGC\,1566, overlaid on the 
greyscale iso-density contour map of the bright blue stellar sources of the galaxy. This sample corresponds to the 92\,per\,cent of the total sample of classified young clusters, indicating that the majority of star cluster formation takes 
place along the arms.  {\em Right panel.} Box plots of the ages and masses of the star clusters. 
The clusters have a mean age $\sim$\,16\,Myr and a mean mass $\sim$\,24$\times 10^3$\,M{\solar}.
The horizontal bars inside the boxes correspond to the medians of the parameters. \label{f:clusagemass}}
\end{figure*}

\subsubsection{Correlations of other parameters with size}\label{s:rho_sz}

In the relations of Fig.\,\ref{f:sizedenscorr} complexes of higher density exhibit progressively smaller sizes and steeper density--size relations. 
This trend is demonstrated by the correlation between the surface stellar mass density $\Sigma_{\star}$ 
of the young stellar structures and their sizes, shown in Fig.\,\ref{f:sizedenscorr}. Points in this figure 
are coloured according to the detection surface density level  (in $\sigma$) of the corresponding structures.
As discussed in the previous section the densities of the detected systems lie between 0.1 and about 1\,M{\solar}\,${\rm pc}^{-2}$. 
Both the plot and the derived statistics indicate an overall dependence of 
stellar mass density on size, the strength of which depends on the density detection level of the structures.
In order to quantify this dependence we apply power-law fits of the form $\displaystyle \Sigma_{\star} \propto S^{-\nu}$,
and estimate the Pearson correlation coefficient, $r_{\rm p}$,  
for each group of detected systems. We tabulate our results in Table\,\ref{tab:raramcorr} (Cols. 4 and 5). 


Low-density complexes (detected at 1 and 2$\sigma$ levels) have almost
 flat exponents ($\nu$ between $0.01$ and $0.06$) and weak correlations ($r_{\rm p} < 0.5$). The flat density--size correlations for these complexes 
agree with the results from the mass--size relations of the previous section, where low-density structures are
found with almost constant surface density. For higher-density structures the power-law slope 
is becoming progressively steeper and the correlation improves. 
For reference, the correlation for the total sample has a moderate negative trend with a correlation coefficient 
$|r_{\rm p}| \simeq 0.55$ (Fig.\,\ref{f:paramcorrel}) and a power-law exponent  $\nu \simeq 0.48$.

The exponents reported in Table\,\ref{tab:raramcorr} satisfy the equality $\kappa = 2 - \nu$ for all sub-samples,
which express the direct relationship between mass, density and size, as expressed by the surface density 
definition ($\Sigma_\star \propto M_{\star} S^{-2}$). The density-size relation can thus be parametrized as
\begin{equation}
\Sigma_\star \propto S^{\kappa-2}. 
\label{fun:msrel}
\end{equation}  
Stellar (volume) density has been proposed as a crucial parameter for the distinction between stellar 
systems of different self-binding strength \citep[][]{1999IAUS..190..410K, gouliermis03} on both theoretical \citep{Bok1934, 
Spitzer1958} and observational \citep{Blaauw1964, ladalada91} grounds. This indicates that stellar density, expressed 
here in terms of observed ``column'' density, is an important intrinsic parameter of stellar groupings. However, it
may not be a fundamental parameter, since it is the derivative of mass and size.
          
Considering that mass and size are basic, independently measured, parameters of the detected structures, 
their relation is fundamental, in the sense that it determines the relations between other derivatives, 
such as density and crossing time, or velocity dispersion and mass (see Fig.\,\ref{f:paramcorrel} for all correlations). 
We can, thus, express all correlations in terms of the mass--size relation and its exponents measured 
for every sub-sample of complexes. For example, following the definitions of the 
crossing time and velocity dispersion of the structures (Sect.\,\ref{s:params}), the relations of these parameters 
with size can be parametrised with the exponent $\kappa$, as in the case of the density--size relation. The derived 
functional forms of the time--size and velocity--size relations are expressed in Table\,\ref{tab:raramcorr}, where the 
corresponding exponents, derived from power-law fits to the data are also given.

\section{Discussion} \label{s:discussion}

In the previous sections we present results on how star formation is organized in a typical grand-design spiral galaxy. 
The stellar complexes of NGC\,1566 are mainly located along its global spiral arms, and they are hierarchically structured 
across the complete observed length-scales range. In this section we discuss three points raised by our study that  
may be important for a comprehensive understanding of global star formation in NGC\,1566. We discuss (1) how the star clusters 
in NGC\,1566 are distributed across the disk of the galaxy in comparison to the stellar complexes, (2) what is 
the fraction of recently-formed stellar mass that is located inside the complexes (most of them along the spiral arms) and in the ``field'', 
and (3) what is the origin of the young stellar populations inside and outside the stellar complexes.

\subsection{Star clusters in the stellar complexes of NGC\,1566} \label{s:disclusters}

Considering that most of the recent star formation, expressed by young stellar over-densities, takes place along the spiral 
arms of the galaxy, an important piece of information would be how many star clusters in the galaxy are also located in the arms.
We cross-correlated the positions of the known star clusters in NGC\,1566 with those of the stellar complexes. The aim was to identify 
the star cluster population that is located within the borders of the stellar complexes identified at the lowest density, 1$\sigma$, level.
The star cluster catalogue is produced by  the LEGUS cluster team through a three-step procedure: 1) Aperture photometry of sources 
that appear to be non-stellar in photometric runs with {\sc Sextractor}, 2) selection of
the best candidates in terms of their concentration index, and 3) final inspection and classification on the multi-band images by eye. The 
detailed description of the procedure is given in Adamo et al. (in preparation).

We found that 480 young clusters, i.e., 70\,per\,cent, of the total young cluster sample are members of the stellar complexes, identified in our study 
at the 1$\sigma$ density level. One hundred and forty-five additional clusters are found outside the borders of the complexes, as defined 
by the 1$\sigma$ isopleths, but within regions that correspond to the average stellar number density (0$\sigma$). These clusters are all located in regions 
between or on the edges of the stellar complexes. Both cluster samples sum to 625 clusters, i.e., 92\,per\,cent of the total sample of classified 
young clusters, which are associated with the star-forming complexes of the galaxy. Their positions are indicated by the black symbols in the map of 
Fig.\,\ref{f:clusagemass} (left panel), showing clearly that the vast majority of star clusters form along the spiral arms. Among the 30 brightest clusters inside stellar complexes six of the youngest ($<$\,50\,Myr) most massive ($>$\,5\,10$^4$\,M{\solar}) clusters are selected by \cite{Wofford2016} 
for a comparative study of various spectral synthesis models against multi-band cluster photometry. 


There are only a few clusters 
concentrated on the far left of the observed field, coinciding with few prominent complexes, which have no obvious relation to the arms. This
over-density of star and cluster formation is not entirely unrelated to the galaxy morphology, as it coincides with bright UV emission (Fig.\,\ref{fig:ngc1566ima}), and 
the corotation ring of the galaxy as defined by, e.g., \cite{Aguero2004}. Box plots of the ages and masses of the clusters in the map are also shown in Fig.\,\ref{f:clusagemass} (right panel). 
The statistics of these parameters indicate that the star clusters in complexes are young and relatively massive.  
The spatial distribution of these young clusters (ages \lsim\,20\,Myr) is not surprising, considering that star clusters are known to be the compact 
parts in the hierarchy of star-forming structures \citep[e.g.][]{efremov89, gouliermis10, Gouliermis2015}, which in the case of NGC\,1566 coagulate mainly along the grand-design symmetric arms of the galaxy. This result is in agreement with models that show that stars and star clusters preferentially form in the spiral arms of galaxies, with their spatial distribution depending on the nature of the arms \citep{DobbsPringle2009}.


\begin{figure*}
\centerline{\includegraphics[clip=true,width=\textwidth]{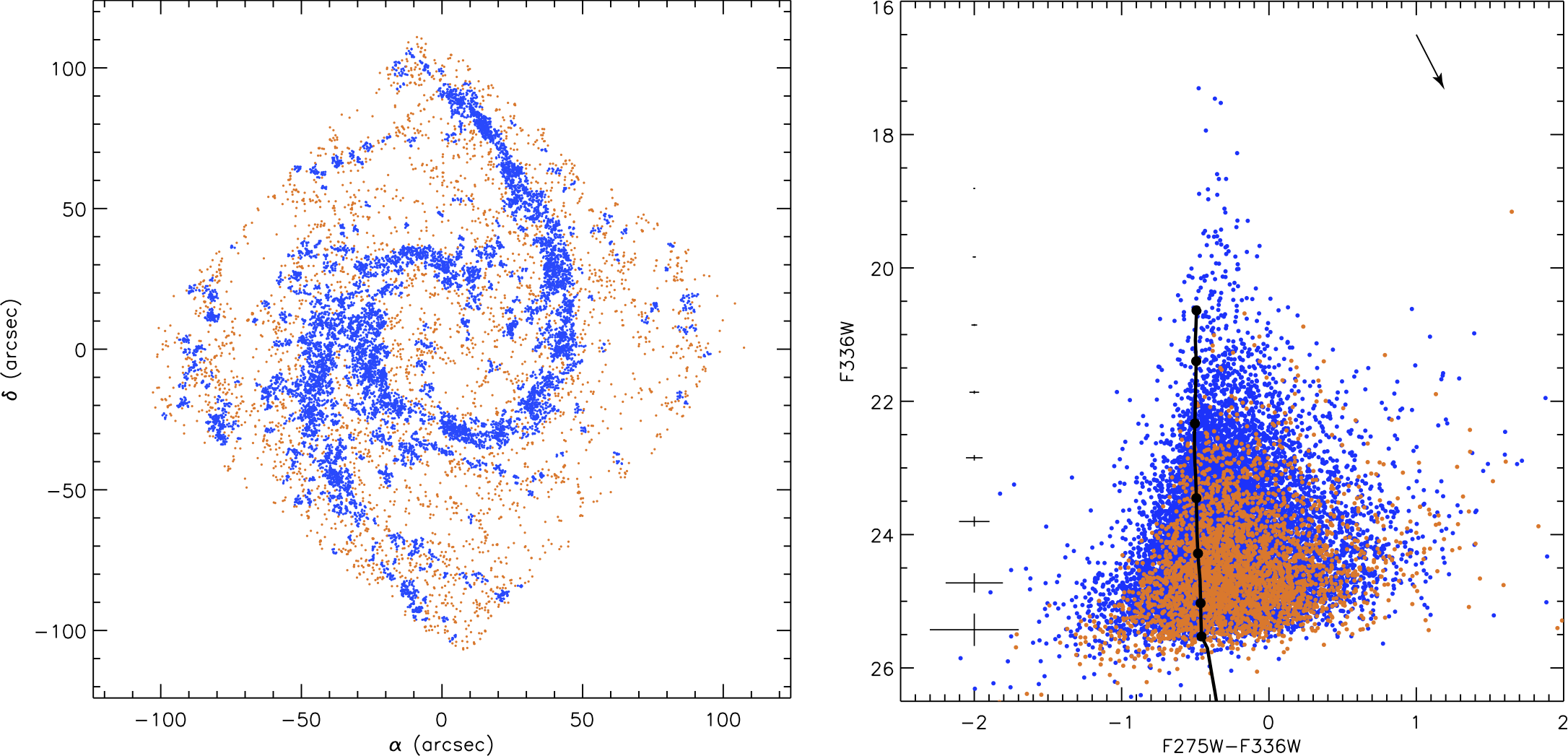}}
\caption{Map (left) and the corresponding CMD (right) of the young blue stellar sources and systems inside the identified stellar 
complexes (blue symbols) and those outside the borders of the complexes (red symbols). While the sources of the 
structures identified as the star-forming complexes of the galaxy are located mostly along the arms, those being more
sparsely distributed (and thus with no significant structure  around them) are located farther from the main arm features.
The latter sub-sample consist of fainter main-sequence stellar sources than the former, indicating that stellar systems more
remote from the arms are in a more advanced stage of their evolution (see Sects.\,\ref{disc:sf_fraction} and \ref{disc:armvsinter}).
The Padova ZAMS for solar metallicity is overlaid in the CMD with the positions for stars with masses 15, 20, 30, 50, 100, 
150 and 300 M$_{\odot}$ indicated by thick dots. An assumed extinction of $A_V =$\,0.55\,mag is indicated by the black arrow, and 
typical photometric uncertainties in both magnitudes and colours are shown on the CMD. \label{f:sysvsfld}}
\end{figure*}

\subsection{The fraction of stellar mass formed in complexes}\label{disc:sf_fraction}

We have shown that young stellar sources in NGC\,1566 form large stellar complexes, most of them located along the spiral arms of the galaxy. 
These stars, however, represent only a certain fraction of the total young stellar population of the galaxy. An important question related to  
star formation across the whole galactic disk is {\sl how much of the recently-formed stellar mass is actually assembled in the star-forming 
complexes, and how much is associated with regions outside these structures (as defined by the 1$\sigma$ isopleths).} In this section we 
answer this question. Recall that the CMD age of our stellar sample, based on the observed UV, U CMD,  is limited to a maximum of $\sim$\,20\,Myr (Sect.\,\ref{s:obs}, Fig.\,\ref{fig:cmds}), and therefore our analysis deals with the most recent star formation in NGC\,1566. 


In the map of Fig.\ref{f:sysvsfld} (left panel) the locations of the stellar members of the complexes (confined within the 1$\sigma$ isopleths) are shown with blue symbols, and those of stellar sources outside the complexes (outside the 1$\sigma$ borders) with red. The CMD positions of the sources in each of the samples are also shown in Fig.\ref{f:sysvsfld} (right panel). 
The corresponding observed mass functions of both samples are constructed as discussed in Appendix\,\ref{s:fieldmf}. The total stellar mass of each of the samples is estimated by extrapolating these MFs (Appendix\,\ref{s:app_smass}, see also Sect.\,\ref{s:param_determ}). The total mass of stars inside the complexes is found $\simeq$\,2.8\,10$^{6}$\,M{\solar}, and that of sources outside the complexes, i.e., in the field, is $\simeq$\,2.4\,10$^{6}$\,M{\solar} (see also Table\,\ref{tab:mfslopes}).   The total stellar mass of the whole blue stellar sample (both inside and outside the complexes boundaries) is determined in terms of extrapolation of its own integrated MF as described in Appendix\,\ref{s:fieldmf}. This mass is $\sim$\,5.3\,10$^{6}$\,M{\solar} (Table\,\ref{tab:mfslopes}). 

The mass values derived above suggest that the stellar mass fraction associated with the star-forming complexes of NGC\,1566 amounts to $\sim$\,53 per\,cent of the total young stellar mass. This  leaves $\sim$\,47 per\,cent of the total stellar mass not being associated with the complexes, at least not directly as we discuss in the next section. The latter fraction is confirmed by the ratio of the stellar mass outside the complexes, independently calculated from the extrapolation of the corresponding MF,  over the mass of the whole blue stellar sample. This ratio equals to $\sim$\,45 per\,cent (see also Table\,\ref{tab:mfslopes}). It should be noted, however, that these fractions may vary. For example, if we consider the total stellar mass confined within the stellar complexes by summing the individual masses of the complexes (derived from their numbers of observed sources multiplied by the determined mass per source of $\sim$\,300\,M{\solar}), this value amounts to $\sim$\,3.5\,10$^6$\,M{\solar} (see also Table\,\ref{tab:structdemo}), which corresponds to $\sim$\,65 per\,cent of the total young stellar mass being formed in stellar complexes. It should be also kept in mind that the stellar masses and the corresponding fractions measured above refer to the observed field-of-view and not the whole extent of the galaxy.


\subsection{Stellar sources in and out of the complexes borders}\label{disc:armvsinter}

Another important question is {\sl if the stellar population associated with regions outside the complexes formed in situ, or if these stellar sources have been removed fast from their natal locations}. There are clear indications that the stellar complexes host the most recently-formed populations. The CMD of Fig.\,\ref{f:sysvsfld} shows the stellar members of the complexes with blue symbols, and 
those sources outside the complexes with red. From this CMD it is seen that  the brightest stars in the ``field'' regions (outside the 1$\sigma$-borders) are much fainter than those in the complexes, with a separation between the populations at $m_{\rm 336} \simeq$\,23\,mag. This may indicate that the bright populations in the field regions are more evolved than those in the complexes, but considering the youthfulness of our stellar sample, they 
should be only marginally older.  On the other hand, the CMD of Fig.\,\ref{f:sysvsfld} includes sources that have been identified as star clusters. Therefore, it seems natural that the brightest sources, being partially clusters, tend to be inside the complexes, while the field includes objects that more likely are individual stars. However, star clusters represent only a small fraction (3 per cent; Sect.\,\ref{s:obs}) of the bright sources in the CMD, and therefore most of these sources are treated as individual stars or unresolved binaries. Under these circumstances, an indicative age-difference that corresponds to the brightness limit between the populations inside and outside the complexes (blue and red symbols in Fig.\,\ref{f:sysvsfld}), as derived from the evolutionary models, is $\sim$\,10\,Myr.

\begin{figure}
\centerline{\includegraphics[clip=true,width=0.475\textwidth]{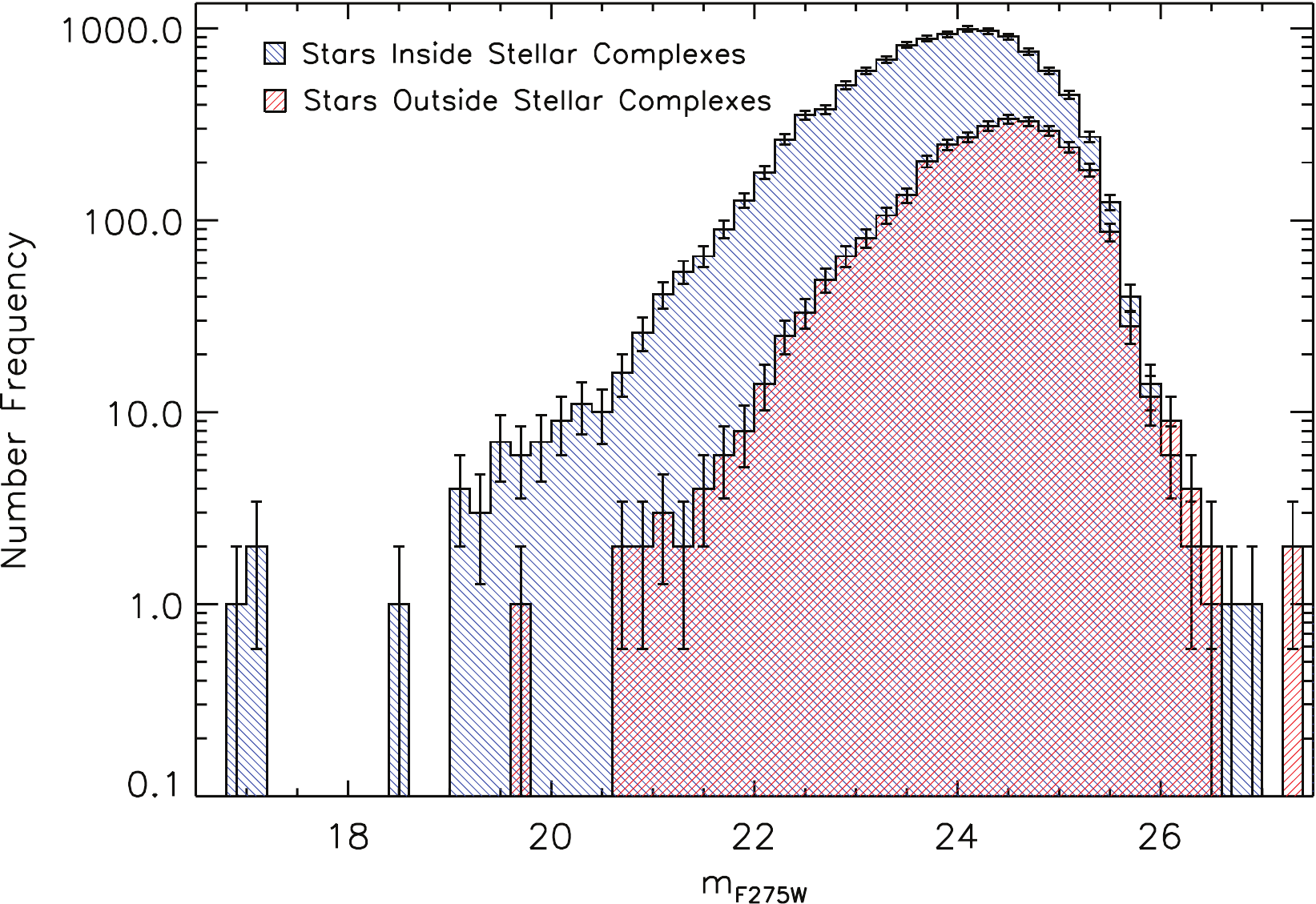}}
\caption{Luminosity functions of blue stellar sources and systems inside the stellar complexes (blue histogram) and those outside the complexes and thus more remote from the main arm features of NGC\,1566 (red histogram). Known star clusters (being among the brightest and reddest sources) are excluded from both samples. The two LFs are identical in the faint regime, but quite different in their bright parts. The lack of stellar sources brighter that $m_{\rm 275}$\,$\sim$\,20.5 
in the ``field'' (outside the complexes) produces a truncated field LF with a steeper slope in comparison to that of stars in the arms.  
This  comparison demonstrates that indeed  the most massive and bright stellar systems are mostly located within the complexes of the galaxy. \label{f:sysvsfldlf}}
\end{figure}

The differences between the two populations are further demonstrated by the luminosity function (LF) of the sources in each sample. In Fig.\,\ref{f:sysvsfldlf} we show the LFs in the F275W (WFC3 UV) filter of both the stellar samples inside and outside the complexes (known clusters are excluded from both LFs). While both catalogues share the same brightness detection limit, set by our photometric sensitivity, their LFs have quite different  shapes in their bright parts, with the LF of the field population being devoid of stars brighter than $m_{\rm 275} \sim$\,20.5\,mag. The statistically significant sample of the field population, however, reaches the limit of $m_{\rm 275} \sim$\,21.5\,mag, which corresponds roughly to stellar mass of $\sim$\,65\,M{\solar}. This stellar mass has a typical lifetime of the order of 10\,Myr, comparable to 
the age limit derived above. On the other hand the stellar LF inside the stellar complexes includes the brightest observed sources that correspond to masses of up to more than $\sim$\,150\,M{\solar}. The LF indicates, thus, that the regions of the complexes host the most recent active star formation events.

The explanation for the populations differences inside and outside the stellar complexes lies in the formation of the spiral structure of NGC\,1566 itself. Grand-design galaxies are the typical examples of spiral structure formation by density waves. In fact, in these galaxies ``large-scale spiral structure {\em is} a density wave''  \citep{BinneyTremaine2008}. According to the density wave scenario, introduced by \cite{LinShu1964}, long-arm spirals are waves that rotate rigidly, where stars and gas enter and leave. 
As molecular clouds move into the density wave they are compressed and the local mass density increases. When it reaches the critical value for Jeans instability, the cloud  will collapse and form stars while being in the arm. 
Moreover, the perpendicular velocity of gas and stars in long-lived spiral arms scales inversely with the density, leading them to spend longer time in the spiral arms than in the interarm regions \citep{2014ApJ...780...32E}. This timescale is expected to be even longer in strongly barred galaxies such as NGC\,1566 \citep{DobbsPringle2013}. Therefore, spiral arms host the youngest most massive stars in the disk. 
These stars, due to their short lifetimes, will die out quickly before they exit the arm. Interarm regions show, thus, a lack of such stars. 

This description is in agreement with the apparent differences in both the stellar content and its distribution between the complexes and the field regions of NGC\,1566. While the latter are not typical counterparts of the interarm regions, they are sparsely distributed by populations somewhat older (or fainter) than those in the arms. On the other hand, the complexes, which include the more ``compact'' young populations, are hierarchical star-forming structures mostly located along the main arms. Nevertheless, it is important to keep in mind that this analysis is based on the blue youngest stellar sources in the galaxy. As such, the populations in the arms of NGC\,1566 are not extremely different from those  outside the arms. The fact that the field regions host stars as massive as $\sim$\,65\,M{\solar} suggests that their populations include stars which are still quite young. Moreover, a closer inspection of the spatial distribution of the population in the field (red symbols in the map of Fig.\,\ref{f:sysvsfld}), shows that this distribution is not entirely unstructured, but follows the general trend of the spiral pattern . 

The separation of the field populations from those inside the complexes is based on the spatial limits set by the 1$\sigma$ isopleths. While these limits specify the borders between statistically significant star-forming complexes and their environments, it does not imply that there is a strict distinction between the structures and their surroundings. On the contrary, the field populations should be considered as the dispersed part of the hierarchical pattern of the stellar arms in NGC\,1566, located at the outskirts of the arms, and eventually populating in the future the interarm regions. This hypothesis is further supported by our finding that even within the same 1$\sigma$ complexes, stellar sources (again excluding the known star clusters) located farther away from the arm ``ridge'' are systematically fainter (and apparently older, similar to the field populations) than those located closer or in it. The evaporation of stellar complexes and the time evolution of galactic-scale stellar distribution have been previously  investigated for the Magellanic Clouds \citep{Gieles2008, Bastian2009}, as well as for the galaxies M\,31 and NGC\,6503 \citep{Gouliermis2015, Gouliermis2015b}, and NGC\,1313 and IC\,2574 \citep{pellerin07, pellerin12}.

Stellar complexes are generally unbound structures and they eventually dissolve through evaporation of their stars, but it is not clear how fast 
this process is\footnote{The crossing time is {\em not} a good estimate for this timescale, since it is only an upper limit based on the observed stellar mass and size. It should also depend on the local environment of the complex. For example, passing-by molecular clouds or shear by the arms 
rotation may increase the kinetic energy of the complex, which will exceed significantly its potential energy and lead to its fast dissolution.}.  
The $\sim$\,10\,Myr difference in age between stars inside the complexes' boundaries and those outside provides a possible minimum timeframe for the brightest young stars to ``escape'' their parental structures. This timescale, however, would be too short for a significant drift from the mid-arm to the mid-interarm regions, because most of the stellar motion in the arms is parallel to the arms. We conclude, thus, that any ``evaporation'' of the complexes must occur to stars, which are already formed close to the borders of the structures. The stellar complexes whose stars are moving out of the arms apparently will be elongated by shear (as is the case, e.g., for few 1$\sigma$ complexes, `emerging' outwards from the eastern arm, as seen in the maps of Figs.\,\ref{fig:contour_map} and \ref{f:sysvsfld}). On the other hand, most of the bright blue stellar sources are possibly formed close to their current locations. We cannot, thus, rule out the possibility that some of the stars outside the complexes were actually formed there (by the density waves) at earlier time, and therefore they are somewhat more evolved.


\section{Summary and Conclusions}\label{s:summa}

We present our clustering analysis of the young blue stellar population detected  
with LEGUS across the grand-design galaxy NGC\,1566. It provides the deepest and 
most complete stellar coverage of the galaxy to date. The application of our  contour-based
clustering technique on the stellar surface density maps of the galaxy  revealed
890 distinct stellar structures, which are the stellar complexes of the galaxy as detected at 
various stellar density (significance) levels. The identified large unbound stellar constellations 
consist of smaller and more compact structures, which themselves ``fragment'' into 
even smaller compact stellar groups. This hierarchical clustering behaviour is quantified by the 
classification of the detected stellar structures into 12 significance levels, in terms of density 
standard deviations ($\sigma$) above the average background density level. The majority of the  
structures 
buildup the spiral arms, down to the 1$\sigma$-level, demonstrating that {\sl star formation 
along the spiral arms of a grand-design galaxy is organized in a hierarchical fashion} 
(Sect.\,\ref{s:detection}, Fig.\,\ref{fig:contour_map}).

We determine several structural parameters for the identified stellar complexes based on their measured sizes and 
stellar sources numbers. The stellar mass, UV brightness, and stellar mass surface density of each complex are 
estimated from these parameters by extrapolating its observed stellar mass spectrum to the sub-solar regime. 
Velocity dispersions (lower limit) and crossing times (upper limit) are determined assuming virial equilibrium for the structures. A strong 
dependence to the density level is found for the average size, average stellar surface density, total stellar mass, total UV brightness, and 
average crossing time of the structures. This indicates that each detection density level corresponds to 
structures with different structural behaviour, with the 1$\sigma$- and 2$\sigma$-level structures corresponding (on average) to the 
most extended, lower-density complexes, which are not well mixed, and those at higher levels being the smaller and more compact structures. 

The size distribution of the complexes peaks around 122\,pc, a length-scale comparable to 
that found for another SAB-type galaxy, the star-forming ring galaxy NGC\,6503, with the same observational material 
and the same technique \citep{Gouliermis2015}. Whether this scale corresponds to a characteristic galactic scale for 
star formation \citep[see, e.g., the discussion in][]{gouliermis11}, or how this scale may depend on galactic environment 
are open issues that should be further investigated with more LEGUS galaxies. 
The size distribution of the stellar complexes at small scales is represented by a log-normal
function. The large-scale side of the distribution shows clear overabundance of structures in 
respect to the Gaussian fit, and is better represented by a power-law. The cumulative size distribution also shows a prominent power-law tail 
of the form $N \propto S^{-1.8 \pm 0.1}$ at large length-scales. {\sl The power-law behaviour of the right-hand part of the size distribution 
indicates a hierarchical mechanism in 
determining the sizes of the large stellar complexes}. 
We explain this part of the distribution with a simple ``hierarchical fragmentation and enrichment'' model, which assumes 
the fragmentation of each ``generation'' of structures into smaller ones and the enrichment of each new generation by 
newly-formed structures in a fashion similar to ``rich-get-richer'' distributions. 

The stellar mass surface density distribution of the identified structures has a bimodal shape, with the 1$\sigma$ structures being well-separated 
from the complexes found in the remaining density levels. Each of the modes are well represented by a log-normal form across the entire observed density ranges with peaks at $\sim$\,0.2 and 0.5\,M{\solar}\,pc$^{-2}$. This implies a clear distinction in density-scales between the low-density 1$\sigma$ structures and their cohort sub-structures. Star formation  would introduce through self-gravity a power-law 
tail in the high-density part of the PDF, which we do not see for the complexes. However, 
this effect would appear at the highest density levels and the length-scales of more compact star-forming clusters and
associations. While the detected complexes are the large structures where stars are forming, our detection limits in both 
size and density do not reach the levels of the compact star-forming centres, which reside inside these complexes. 
Therefore, we do not observe any power-law tail in the densities PDF of the complexes. On the other hand 
log-normal density PDFs, like that of the complexes, are characteristic for supersonic non-gravitating turbulent gas.
If we assume that the structures of young stars do inherit their morphology from their parental ISM, then the observed density 
PDF is a clear indication that {\sl the formation of the identified stellar complexes is driven by the large-scale turbulence
in the galactic disk}.


There are strong correlations between the structural parameters of the identified complexes. The mass--size relation of the 
complexes shows a power-law shape, determined through linear regression in the log-log parameters space. 
The exponent of this relation is found to depend on the significance level of the structures, with that for the highest-level 
structures being equal to $\sim$\,1.5, and that for the lowest-level structures being almost equal to 2. The latter exponent, 
which is found for both 1$\sigma$- and 2$\sigma$-level structures, indicates that the stellar surface density of these structures 
is independent of their size, i.e., all structures have the same surface density. On the other hand, exponents of the mass--size relation 
smaller than 2, found for the higher significance structures, indicate that these complexes can differ significantly in surface density at 
given scale. The mass--size relation exponent derived for the whole sample, i.e. for all detected 
structures at various levels, is $\simeq$\,1.52. Considering that fractional exponents are found for the mass--size 
relations of fractal GMCs \citep[e.g.,][]{elmefalga96, Sanchez2005}, {\sl the mass--size relation of the stellar 
complexes in NGC\,1566 provides evidence of their self-similar morphology}. Dependences of the power-law exponents on the 
significance level are found also for the correlations of surface density, crossing time and velocity dispersion with size. We 
thus identify size as the fundamental structural parameter of the complexes.

A significant fraction, specifically 92\,per\,cent, of the known young ($\leq$\,100\,Myr) stellar 
clusters in NGC\,1566 are located inside or in the direct vicinity of the identified stellar complexes. Considering that the majority of the 
identified complexes are located along the arms of NGC\,1566, this finding 
confirms galactic-scale star formation models, according to which clusters preferentially form in the spiral arms. About 50 to 65 per\,cent of the total blue stellar mass of the galaxy (within the observed field-of-view) 
is located inside the identified stellar complexes (within the borders of the 1$\sigma$ structures), with the remaining stellar mass occupying their fields, still following the spiral arms morphology. This ``field'' population is fainter and more sparsely distributed  that that in the complexes. The differences between the populations inside and outside the complexes allow also for different mass functions, with that of the field population being  significantly steeper than that of stellar sources inside the complexes (Appendix\,\ref{s:fieldmf}). 

{The CMD positions of the stars suggest an {\sl age-difference between the stellar sources 
inside and outside the complexes of the order of $\sim$\,10\,Myr}, determined from the stellar upper brightness limit on the main sequence. This age-difference may provide a timescale for stars to move out of their natal structures. However, considering that most of the observed young stars are formed close to their current positions, this ``evaporation'' concerns stars formed already close to the borders of the complexes. Moreover, we cannot rule out the possibility that some of the inter-complex bright blue stars were actually formed there. For example, there could be GMCs and star formation occurring outside the complexes, in accordance to our ``enrichment'' scenario. So, there could be both enrichment (in situ formation) and drift. In any case, the populations, which lay outside the 
stellar complexes but still in their environments, should be considered as the dispersed part of the hierarchical stellar distribution along the arms of NGC\,1566.

Our study shows that most of the very young stellar mass in NGC\,1566 is concentrated into spiral arms. This finding is in agreement with the distribution of H\,{\sc ii} regions across the main body of the galaxy \citep{1982A&A...114....7C}. We have also shown that young stars and stellar systems are assembled in large stellar complexes that buildup the stellar spiral arms of the galaxy in a {\sl hierarchical} fashion. Some of these, particularly in the western arm, form a structure that resembles ``beads on a string'', which is usually associated with large scale gravitational instabilities along the arms. Our findings agree, thus, with the hypothesis of turbulence-driven hierarchical star formation across galactic scales as has been previously 
observed with LEGUS from UV images \citep{2elmegreen14}, resolved stellar populations \citep{Gouliermis2015}, and stars 
clusters \citep{Grasha2015} in various nearby galaxies. These studies provide evidence that galaxy-wide star formation is organized by large-scale gravitational processes in a pattern analogous to the turbulent self-similar galactic ISM structure. 

\section*{Acknowledgments}
D.A.G. kindly acknowledges financial support by the German Research Foundation (DFG) through program GO\,1659/3-2. He also would like to thank S. Hony for the long discussions (and debates) on star formation and the derived mass function. 
This research has made use of the NASA/IPAC Extragalactic Database (NED), which is operated by the Jet Propulsion Laboratory, 
California Institute of Technology, under contract with the National Aeronautics and Space Administration.
This research has made use of the SIMBAD database, operated at CDS, Strasbourg, France. 
Based on observations made with the NASA/ESA {\sl Hubble Space Telescope}, obtained from the data archive at the 
Space Telescope Science Institute (STScI). STScI is operated by the Association of Universities for Research in Astronomy, Inc.\ under 
NASA contract NAS 5-26555. These observations are associated with program GO-13364. Support for Program 13364 was provided by NASA 
through grants from STScI. This research made use of the TOPCAT\footnote{TOPCAT is available at the permalink \href{http://www.starlink.ac.uk/topcat/}{http://www.starlink.ac.uk/topcat/ }} application \citep{topcat2005}, the R environment for statistical computing 
and graphics\footnote{The R Project for Statistical Computing: \href{https://www.R-project.org/}{https://www.R-project.org/ }} \citep{Rpackage2015}, 
and NASA Astrophysics Data System (ADS) bibliographic services\footnote{Accessible at \href{http://adswww.harvard.edu/}{http://adswww.harvard.edu/} 
and \href{http://cdsads.u-strasbg.fr/}{http://cdsads.u-strasbg.fr/}}.


\appendix

 \section{The Integrated Young Stellar Mass Function}\label{s:fieldmf}

In order to explore how the integrated mass function of young stars changes across the galactic disk of NGC\,1566  we select three samples from the complete observed young stellar inventory, and construct the corresponding mass spectra. These samples are: 1) The total observed blue stellar catalogue, 2) all blue sources that belong to stellar complexes (encompassed within the 1$\sigma$ isopleths), and 3) all remaining stellar sources outside the complexes, i.e., those located in the {\em disk field}. A rough determination of the mass of every stellar source in each of these samples is made from its position in the blue CMD (Fig.\,\ref{fig:cmds}) by interpolating its magnitude and colour to the corresponding ZAMS mass for a fixed distance modulus and constant extinction (both specified in Sect.\,\ref{s:obs}). We construct the stellar mass spectrum of each stellar sample by counting the stars in linear mass bins  down to the detection limit of $\sim$\,20\,M{\solar}. 
The mass spectrum has the form:
\begin{equation}
f(m) \equiv \frac{dN(m)}{dm} \propto m^{\gamma}, 
\end{equation}               
where $\gamma$ is the exponent of the spectrum, which for a power-law is independent of mass. We refer to this mass spectrum as 
{\em mass function} (MF) throughout the paper\footnote{The stellar mass function, $\xi(\log m) \propto m^{\Gamma}$, is by definition 
constructed by counting stars in logarithmic base 10 
bins and not in linear bins as the mass spectrum \citep[e.g.,][]{Gouliermis2006mf}. They are two different functions occasionally confused in the 
literature. For a power-law MF its slope $\Gamma$ relates to $\gamma$ as $\Gamma = \gamma +1$ \citep[e.g.,][]{Scalo1986}. Therefore, a
\cite{Salpeter1955} MF of slope $\Gamma = -$1.35 corresponds to a mass spectrum of exponent $\gamma = -$2.35.}. 

The MF slopes derived for each stellar sample are given in Table\,\ref{tab:mfslopes}. The corresponding MFs are plotted in Fig.\,\ref{f:obsmfs}. A notable observation is that both the MF of the whole sample and that of the populations in the complexes have similar slopes, both being compatible with \cite{Salpeter1955} and \cite{Kroupa2002} MFs in the same mass range. Another interesting result from the slopes of Table\,\ref{tab:mfslopes} and Fig.\,\ref{f:obsmfs} is that  the 
MF of the field  is steeper than both the MF of the whole blue stellar sample, and the MF of the populations in star-forming complexes. This difference follows the change of the luminosity function between the populations inside the complexes and those outside their borders (Sect.\,\ref{disc:sf_fraction}). The LF of the field is steeper than that in the complexes and devoid of the more massive and brighter stellar sources. 

\begin{table}
\centering
\caption{PDMF slopes, number of detected stellar sources and total stellar mass derived from extrapolation of the corresponding observed MFs to the low-mass regime, for the three global subsamples of the blue stellar population of NGC\,1566: (i) The whole observed 
 sample, (ii) sources encompassed within the 1$\sigma$ isopleths, i.e., belonging to stellar complexes, and (iii) sources located outside the 
 complexes borders (in the surrounding field).}
\label{tab:mfslopes}
\begin{tabular}{r c r c}
\hline
 &
{MF Slope}  &
\multicolumn{1}{c}{$N_\star$} &
Total Mass \\
& 
($\gamma$)  &
\multicolumn{1}{c}{($10^3$)} &
($10^6$\,M{\solar}) \\
\hline 
                     Whole Sample & $-2.35 \pm 0.19$ & 14.94 & 5.3 \\ 
        Sources in Complexes & $-2.22 \pm 0.21$ & 11.74 & 2.8 \\
                       Field Sources & $-2.99 \pm 0.12$ & 3.20 &  2.4\\
\hline
\end{tabular}
\end{table}

Differences in the MFs between field and clustered stellar populations are previously reported in the literature. For example, the IMF of field massive stars is found to be much steeper than that of star-forming clusters and associations in the Magellanic Clouds \citep[][and references therein]{Massey2003}. In general, star-forming systems have a preference to top-heavy MFs, while the field shows mostly bottom-heavy MFs \citep[e.g.,][]{Gouliermis2002}. The variation of integrated galactic MF slopes from bottom- to top-heavy has been explained by the differences in galaxy-wide star formation rates \citep[e.g.,][]{Weidner2010, Kroupa2014}, as the areal SFR inside the young stellar systems is higher than away from them. While this scenario can explain the observed variations in the integrated blue MF of NGC\,1566, one should keep in mind that these MFs correspond to the {\em present-day mass function} (PDMF), and not to the {\em initial mass function} (IMF), at least for some of the considered populations.  Some of these PDMFs are thus the products of  both star formation {\em and} stellar evolution, as well as dynamics, since dynamical effects can become significant in massive compact systems even at very young ages. In the following section the total mass of young stellar sources in NGC\,1566 is evaluated through the extrapolation of the observed MFs.

\begin{figure}
\centerline{\includegraphics[clip=true,width=0.475\textwidth]{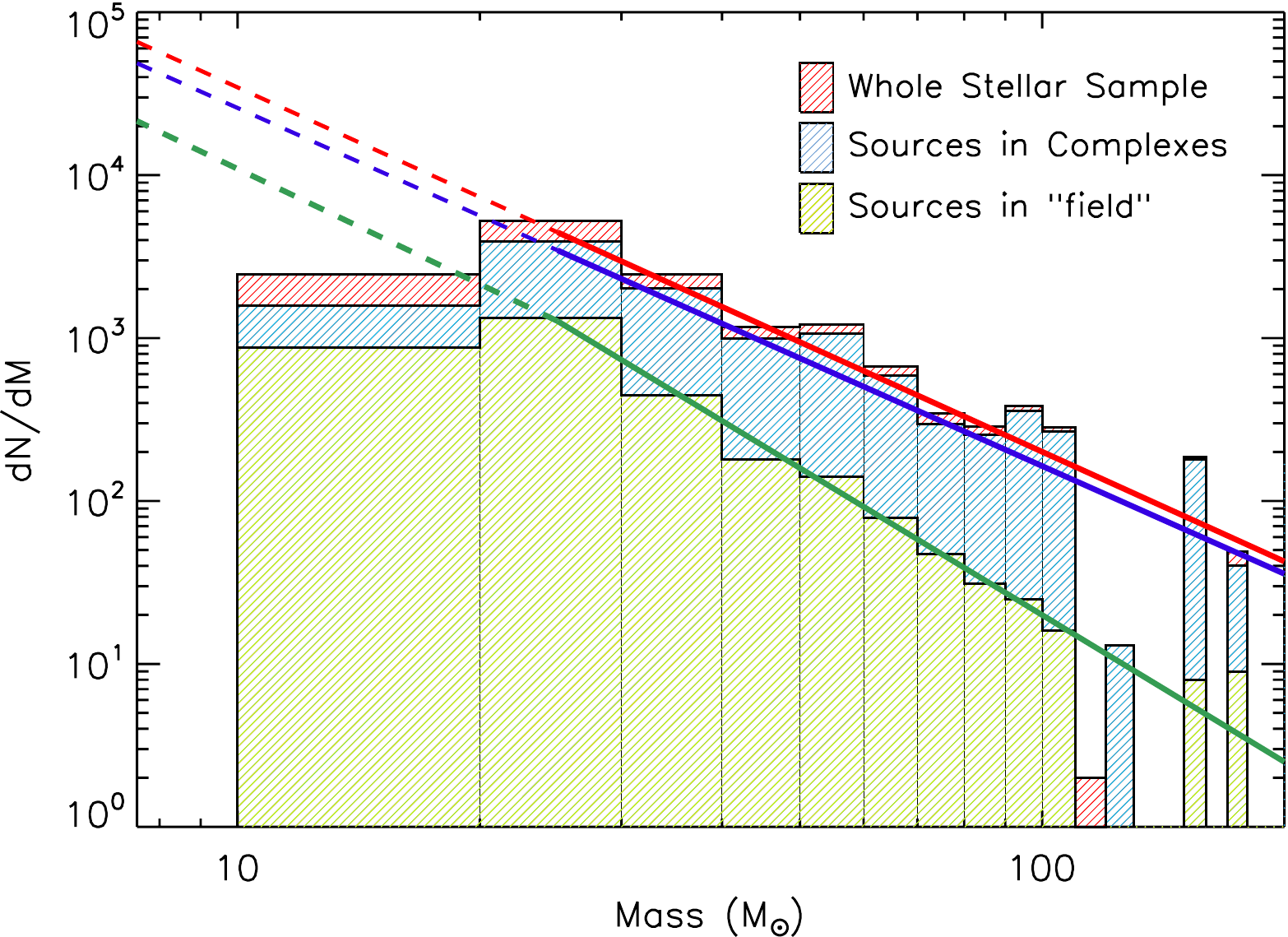}}
\caption{The PDMF of the three selected global blue stellar samples in NGC\,1566: (1) the whole
sample (red histogram), (2) the sample of sources in stellar complexes (blue histogram) and (3) the
sample of sources outside the complexes, i.e., in the ``field'' region (green histogram). The slopes
of the MFs are depicted with their fitted lines. The MF slopes of the whole sample and the sources in 
the star-forming complexes are practically identical, while that of the field population is steeper,
due the effect of the star formation history to the PDMF (as discussed in Appendix\,\ref{s:app_smass}). The dashed lines represent the 
extrapolation applied to these MFs. The MFs of the whole sample and of the populations in the complexes were extrapolated following the 
slope of their observed high-mass regime (as measured down to $\sim$\,20\,M{\solar}). These slopes were applied for extrapolation 
down to $\sim$\,0.5\,M{\solar}. The field MF was extrapolated following 
its measured slope ($\sim-$3) for stars down to $\sim$\,20\,M{\solar}. Then a Salpeter IMF with slope $-$2.35 
was assumed for stars down to $\sim$\,0.5\,M{\solar}. For sub-solar masses down to 0.1\,M{\solar} the Kroupa MF slope 
$-$1.3 was applied for all three MFs. \label{f:obsmfs}}
\end{figure}


\subsection{Total young stellar mass in NGC\,1566 from MF extrapolation}\label{s:app_smass}

Considering that the MF of high-mass stellar sources in all star-forming complexes has the slope comparable to the canonical IMF, it appears that this is the stellar IMF of the young populations, mostly located in the spiral arms. However, the field regions (sources outside the borders of the complexes) show a steeper stellar MF, indicating that it is affected by stellar evolution. Specifically, stars with lifetimes older than the observed age of the blue population\footnote{The older observed age for the blue population of NGC\,1566 is about 20\,Myr (see Sect.\,\ref{s:obs}, and Fig.\,\ref{fig:cmds}).} will be present with their full IMF, but stars with shorter lifetimes will be present in a PDMF steeper than the IMF, because in that time they have been evolving and vanishing \citep{elmegreen-scalo-2006}.

\begin{figure}
\centerline{\includegraphics[clip=true,width=0.475\textwidth]{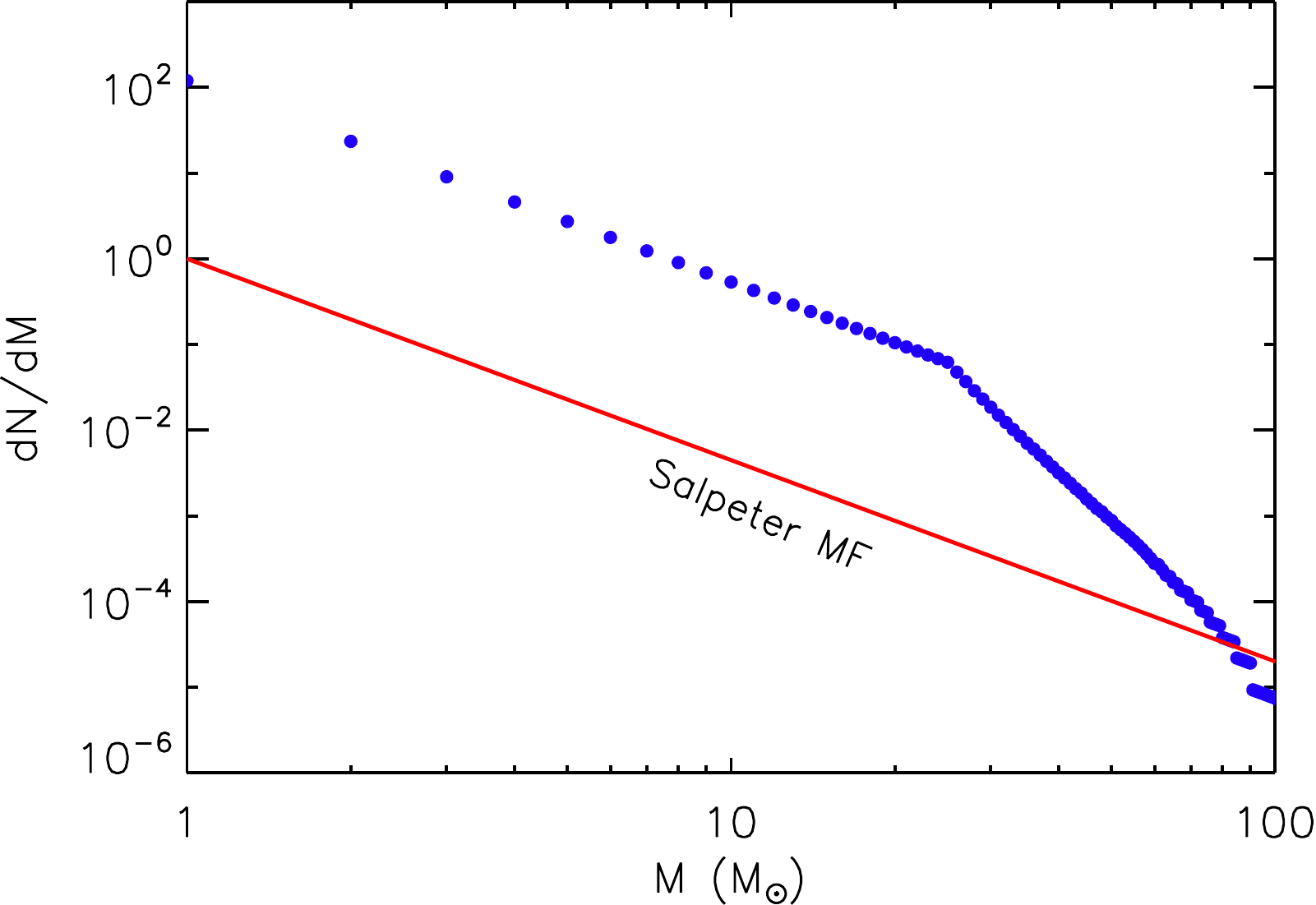}}
\caption{The effect of the star formation history on the PDMF.  The PDMF of a stellar population with a Salpeter IMF, 
which started forming $\sim$\,190\,Myr ago is plotted with blue symbols. This artificial PDMF follows 
specific prescriptions for stellar lifetime over mass and SFR over time (see Appendix\,\ref{s:app_smass}). 
The high-mass regime of the PDMF shows a deficiency of stars with lifetimes shorter than 190\,Myr 
(stars with $M>$\,25\,M{\solar}), which are therefore being progressively vanished. A Salpeter IMF is shown
with a red line for reference.
\label{fig:imfpdmf}}
\end{figure}

Let us assume a galaxy where star formation started $T$ time ago, with $R(t)$ being the change of the star formation rate (SFR) with time $t$ (normally increasing toward the past), and $\tau(M)$ the lifetime of a star of mass $M$. Then the integrated PDMF, $n(M)dM$, of the galaxy is given by the integral
\begin{equation}
\int_{t=0}^{\min(T, \tau(M))} R(t)\, n(M)\, dt = n(M) \int_{t=0}^{\min(T, \tau(M))} R(t)\, dt.
\label{eq:pdmf}
\end{equation}                                                       
The PDMF is thus given by the IMF times the proportion of stars formed in the past time equal to the age of a star of mass $M$.
As a consequence, the high-mass PDMF becomes steeper than the IMF, as an effect of the star formation history. This is visualized in Fig.\,\ref{fig:imfpdmf}, where the results of Eq.\,(\ref{eq:pdmf}) are shown for star formation starting at $T\sim 190$\,Myr ago, a typical dependence of stellar lifetime on mass $\tau(M) = 3(M/100)^{-3}$\,Myr, and an exponentially declining SFR with time $R(t) \propto \exp(-t/T)$, assuming a Salpeter IMF (mass spectrum slope $-$2.35). The PDMF is steeper than the IMF for the mass range 
down to the mass limit, $M_{\rm IMF}$, where $\tau (M_{\rm IMF}) = T$. For smaller masses the PDMF has the IMF slope, since all stellar masses correspond to their initial values.

Taking the above into account, we extrapolate the observed {\sl young field} PDMF of NGC\,1566 to the low-mass regime considering evolutionary effects. Therefore, we assume that this MF follows the same measured slope only for the observed mass range, down to $\sim$\,20\,M{\solar}, while for smaller masses it becomes similar to the typical IMF with slope $-$2.35 down to the stellar mass of 0.5\,M{\solar}. On the other hand, both the MF of the stellar complexes population, and that of the whole sample have typical Salpeter IMF slopes, suggesting that they are not significantly affected by evolutionary effects. We extrapolate, thus, these MFs  down to the limit of 0.5\,M{\solar}, assuming that they follow their measured  slopes ($-$2.22 and $-$2.35 respectively). All three MFs are then extrapolated to the stellar mass limit of 0.1\,M{\solar}, assuming the Kroupa IMF with a slope $-$1.3. We derive the complete total stellar mass of each population based on the extrapolated MFs. 

The corresponding measured stellar masses are given in Table\,\ref{tab:mfslopes}. From these values it can be seen that the sum of the total stellar mass encompassed within the stellar complexes and that of the field stellar sources outside the complexes (both calculated independently from the extrapolation of their own MFs) is almost identical to the total stellar mass derived from the independent MF extrapolation of the whole observed blue stellar population in NGC\,1566. This provides confidence to our extrapolation of the field MF by assuming evolutionary effects. 
The measured total young stellar mass,  $M_{\rm tot} \simeq$\,5\,10$^6$\,M{\solar}, divided by the total number of observed stellar sources in the galaxy provides an estimate of the average 
{\sl true} stellar mass per detected source. This mass, which amounts to $\sim$\,350\,M{\solar}, corresponds to the whole blue stellar sample. If we consider only the number of stars included within the 1$\sigma$ isopleths (i.e., the stellar complexes members) and their corresponding total mass ($\simeq$\,3\,10$^6$\,M{\solar}), the mass per sources equals to $\sim$\,250\,M{\solar}. We determine the total stellar mass of individual stellar complexes assuming a mass per detected sources that corresponds to the average of the two measurements, i.e., to $\sim$\,300\,M{\solar}. The total mass of each complex was 
evaluated from the number of its stellar members multiplied by this mass (see Sect.\,\ref{s:param_determ}).

\subsection{The Star Formation Rate in NGC\,1566}\label{s:app_sfr} 

The {\sl star formation rate} (SFR) is an exceptionally useful parameter in understanding star formation across galactic scales. It is, however, the most contradictive in its calculation from stellar samples alone. A naive determination of the SFR from the total young stellar mass of the galaxy assumes a fixed age $T$ for all stars in the sample, suggesting a simple scaling relation between $\displaystyle M_{\star}$ and SFR:
\begin{equation}
\displaystyle {\rm SFR} = \frac{M_{\star}\,{\rm (M}{\solar}{\rm )}}{T\,\rm (Myr)}.
\label{eq:sfr}
\end{equation}
However, this conversion of young massive stars to SFR assumes that the latter is constant for all times into the past, which is not always the case 
for spiral galaxies like NGC\,1566. In these galaxies  SFR is also assumed to be higher in the arms and to change as a function of arm phase 
 \citep[e.g.,][]{Knapen1992, Knapen1996}, making the determination of the SFR and the SFR surface density ($\displaystyle \Sigma_{\rm SFR}$) 
 from star counts quite complicated.  Nevertheless, the application of  the simple conversion of Eq.\,(\ref{eq:sfr}), assuming an age for the stellar 
 population of 20\,Myr, results to a constant SFR of $\sim$ 0.27\,M{\solar}yr$^{-1}$. 
 
 In order to evaluate this result we performed a set of simple population 
 synthesis simulations tailored to our observed data for NGC\,1566. Specifically, we do not consider binaries, photometric errors, and incompleteness, and we assume that 
 the populations follow a  Kroupa IMF with a constant SFH. The simulations were normalized to the observed number of stars in NGC1566 
 brighter than $m_{\rm 336}=24$, which for a constant SFH between 0 and 20 Myr does not correspond to a single mass but to a distribution 
 between 12 and 26\,M{\solar}. Repeated simulations for different timescales of constant SFH (including very narrow timescales, which result to single-mass stars) derived SFRs between\,$\sim$\,0.2 and 0.3\,M{\solar}yr$^{-1}$, consistent with our estimation based on the extrapolated total young stellar mass\footnote{These SFRs are lower but still comparable to the (dust corrected) SFR derived from integrated H$_{\alpha}$, UV, and IR fluxes by \cite{ZhouZ2015}, which, adjusted to our distance for the galaxy, amounts to 1.5\,M{\solar}\,yr$^{-1}$.}. This agreement provides additional confidence to our measurement  of the total mass in young stars in NGC\,1566, and the subsequent determination of the mass of individual complexes according to their observed stellar numbers. 

\section{The effect of the distance of NGC\,1566}\label{s:dist_effect}

In this study we assume a distance of $\sim$\,10\,Mpc for NGC\,1566, in agreement to NED average for this galaxy, and our 
isochrone fitting of the RGB tip in the HST optical (F555W, F814W) CMD. In this section we shortly discuss the effect of a larger distance 
for the galaxy to our results, and we show that our main findings are not sensitive to the assumed distance. Specifically, assuming the largest estimated distance for NGC\,1566 
($\sim$\,20\,Mpc) would lead the determined length-scales to increase by a factor of $\sim$\,2, 
introducing a shift in the size distribution of the identified structures (Sect.\,\ref{s:size}) but not changing its shape.
The surface stellar mass density would experience a small decrease (due to the systematic increase in both size and mass), 
but its distribution again would not have a different shape (Sect.\,\ref{s:surf_dens_distr}). Likewise, a larger distance modulus 
(by \gsim\,1.5\,mag) would increase the calculated stellar masses by a factor of at least \gsim\,3.5 (the mass--luminosity 
relation is shown in Fig.\,\ref{f:masslight}). This would introduce a new intercept in the log-log correlation of mass with size without altering its slope 
(Sect.\,\ref{s:param_correl}), and subsequent new normalisation in the correlations of size with other mass-derived parameters, 
again without affecting the exponents of their power-law relations (Table\,\ref{tab:raramcorr}). The total young stellar 
mass determined by extrapolation of the mass function (Appendix\,\ref{s:fieldmf}, see also Sect.\,\ref{s:param_determ}) will be also affected by 
a larger distance for NGC\,1566, but the fractions of stellar mass formed inside and outside stellar complexes (discussed in Sect.\,\ref{s:discussion}) will not, since all mass estimates will be systematically increased. The adopted distance for NGC\,1566 of $\sim$\,10\,Mpc 
was selected with these uncertainties in mind.

\end{document}